\documentclass{aa}

\usepackage{txfonts}
\usepackage{graphicx}
\usepackage{array}
\usepackage{booktabs}
\usepackage[colorlinks=true,allcolors=blue]{hyperref}

\newcommand{\ffdeg}{\mbox{\ensuremath{.\!\!\degr}}}

\newcommand{\ffarcs}{\mbox{\ensuremath{.\!\!^{\prime\prime}}}}

\newcolumntype{L}[1]{>{\raggedright\let\newline\\\arraybackslash\hspace{0pt}}m{#1}}
\newcolumntype{C}[1]{>{\centering\let\newline\\\arraybackslash\hspace{0pt}}m{#1}}

\begin{document}

\title{PynPoint: a modular pipeline architecture for processing and analysis of high-contrast imaging data\thanks{Based on observations collected at the European Southern Observatory, Chile, ESO No. 60.A-9800(J), 084.C-0739(A), and 090.C-0653(D)}\fnmsep\thanks{PynPoint is available at \url{https://github.com/PynPoint/PynPoint} under the GNU General Public License v3.}}

\author{
T.~Stolker\inst{1}
\and M.\,J.~Bonse\inst{1}
\and S.\,P.~Quanz\inst{1}\thanks{National Center of Competence in Research "PlanetS" (\url{http://nccr-planets.ch})}
\and A.~Amara\inst{1}
\and G.~Cugno\inst{1}
\and A.\,J.~Bohn\inst{2}
\and A.~Boehle\inst{1}
}

\institute{
Institute for Particle Physics and Astrophysics, ETH Zurich, Wolfgang-Pauli-Strasse 27, 8093 Zurich, Switzerland\\
\email{tomas.stolker@phys.ethz.ch}
\and Leiden Observatory, Leiden University, P.O. Box 9513, 2300 RA Leiden, The Netherlands
}

\date{Received ?; accepted ?}

\abstract
{The direct detection and characterization of planetary and substellar companions at small angular separations is a rapidly advancing field. Dedicated high-contrast imaging instruments deliver unprecedented sensitivity, enabling detailed insights into the atmospheres of young low-mass companions. In addition, improvements in data reduction and point spread function(PSF)-subtraction algorithms are equally relevant for maximizing the scientific yield, both from new and archival data sets.}
{We aim at developing a generic and modular data-reduction pipeline for processing and analysis of high-contrast imaging data obtained with pupil-stabilized observations. The package should be scalable and robust for future implementations and particularly suitable for the 3--5~$\mu$m wavelength range where typically thousands of frames have to be processed and an accurate subtraction of the thermal background emission is critical.}
{PynPoint is written in Python~2.7 and applies various image-processing techniques, as well as statistical tools for analyzing the data, building on open-source Python packages. The current version of PynPoint has evolved from an earlier version that was developed as a PSF-subtraction tool based on principal component analysis (PCA).}
{The architecture of PynPoint has been redesigned with the core functionalities decoupled from the pipeline modules. Modules have been implemented for dedicated processing and analysis steps, including background subtraction, frame registration, PSF subtraction, photometric and astrometric measurements, and estimation of detection limits. The pipeline package enables end-to-end data reduction of pupil-stabilized data and supports classical dithering and coronagraphic data sets. As an example, we processed archival VLT/NACO $L'$ and $M'$ data of $\beta$~Pic~b and reassessed the brightness and position of the planet with a Markov chain Monte Carlo (MCMC) analysis; we also provide a derivation of the photometric error budget.}
{}


\maketitle

\section{Introduction}\label{sec:introduction}

High-contrast imaging is a powerful technique to study the population of planetary and substellar companions at orbital radii beyond $\sim$5--10~au \citep[e.g.,][]{oppenheimer2009,bowler2016}. Although the occurrence rate of gas giant exoplanets on long-period orbits is low, as reflected by the numerous nondetections of large-scale surveys \citep[e.g.,][]{biller2013,brandt2014,galicher2016}, directly imaged planets are key targets for atmospheric characterization \citep[e.g.,][]{barman2015,morzinski2015,rajan2017}. The direct imaging technique is biased towards high-temperature gas giant planets at an early age ($\lesssim 100$~Myr) because contraction of their atmospheric envelope makes these objects bright at near-infrared (NIR) wavelengths \citep[e.g.,][]{burrows1997,marley2007}. The family of directly detected exoplanets includes HR~8799~b, c, d, e \citep{marois2008,marois2010b}, $\beta$~Pic~b \citep{lagrange2009,lagrange2010}, HD~95085~b \citep{rameau2013a,rameau2013b}, 51~Eri~b \citep{macintosh2015}, and HIP~65426~b \citep{chauvin2017}. These objects have masses below the deuterium burning limit, orbit at separations of several tens of astronomical units, and are low in mass compared to their host star ($M_\mathrm{planet}/M_\mathrm{star} \lesssim10^{-3}$), making them particularly interesting regarding their formation and evolutionary pathway.

Understanding the physical and chemical characteristics of long-period giant planets and constraining their orbital architectures are two of the main goals for direct imaging campaigns. During recent years, high-precision spectrophotometric characterization has become possible, especially with the advent of extreme adaptive optics (XAO)-assisted, high-contrast imaging instruments such as VLT/SPHERE \citep{beuzit2008} and Gemini/GPI \citep{macintosh2008}. Dedicated instrumentation and differential observing techniques are key elements that drive the field of direct imaging forward in parallel with improvements in data-processing algorithms.

\begin{figure*}
\centering
\includegraphics[width=\textwidth]{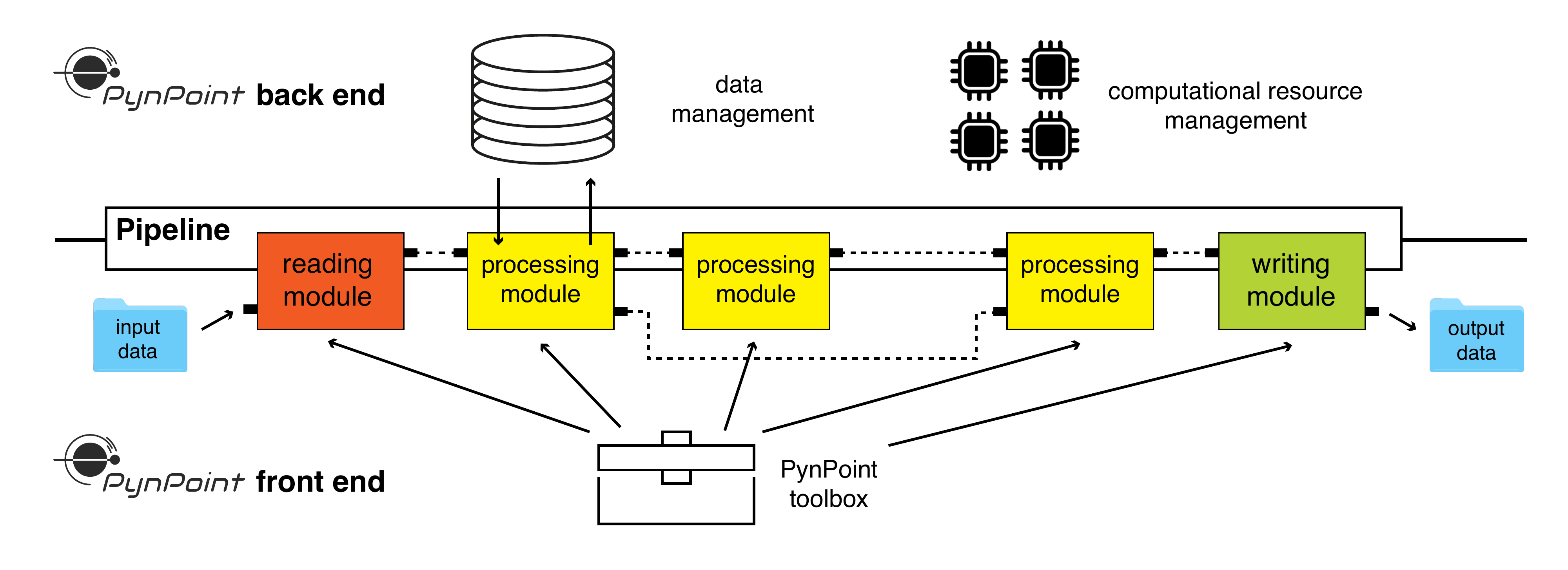}
\caption{Schematic overview of the software design with the separation of the front-end and back-end functionalities. PynPoint offers a simple front end which can be used to define a sequence of pipeline modules. Management of the data and the computational resources (i.e., multiprocessing and memory usage) is handled by the back-end of PynPoint. Reading, processing, and writing modules are attached to the pipeline and sequentially executed while results are stored in the central database. The architecture allows the user to easily rerun pipeline modules and evaluate the results at various stages of the data reduction.\label{fig:pipeline}}
\end{figure*}

Angular differential imaging (ADI) is a particularly powerful observing technique because it allows the data itself to be used as reference for the PSF subtraction as the field is rotating with respect to the telescope pupil \citep{marois2006}. Over the last decade, various post-processing and/or detection algorithms have been introduced that exploit the parallactic rotation of the data to subtract the stellar halo and quasi-static speckles. For example, locally optimized combination of images \citep[LOCI;][]{lafreniere2007} is a method which is based on a least-squares minimization of residual speckle noise. Several derivatives of LOCI have been proposed such as d-LOCI \citep{pueyo2012}, TLOCI \citep{marois2014}, and MLOCI \citep{wahhaj2015}. Other methods include principal component analysis \citep[PCA;][]{amara2012,soummer2012}, angular differential optimal exoplanet detection algorithm \citep[ANDROMEDA;][]{cantalloube2015}, local decomposition into low-rank, sparse, and Gaussian noise components \citep[LLSG;][]{gomez-gonzalez2017}, and supervised machine learning \citep{gomez-gonzalez2018}.

Post-processing algorithms have been implemented in instrument-specific pipelines such as the data cube extraction and speckle suppression pipelines for Project 1640 \citep{zimmerman2011,crepp2011}, the open-source ACORNS-ADI pipeline of the Subaru/HiCIAO SEEDS survey \citep{brandt2013}, the automated data processing architecture of the Gemini Planet Imager Exoplanet Survey \citep[GPIES Data Cruncher;][]{wang2018}, and the SPHERE speckle calibration (SpeCal) tool of the SPHERE consortium \citep{galicher2018}. These pipelines allow for end-to-end processing of high-contrast data from specific instruments but are not publicly available in most cases. There are also a few generic pipelines, including SOSIE \citep{marois2010a} and GRAPHIC \citep{hagelberg2016}, which are suitable for multiple instruments but are (currently) not publicly available. Most recently,
 \citet{gomez-gonzalez2017}   presented the Vortex Image Processing (VIP) package,  an open-source library of Python functions dedicated to high-contrast imaging data.

In this paper, we present the new architecture of PynPoint: a generic, open-source pipeline for processing and analysis of high-contrast imaging data obtained with ADI. PynPoint was originally developed as a PSF-subtraction tool with PCA \citep{amara2015} while the new Python package provides a pipeline for end-to-end data reduction, including various analysis tools. The new architecture has a modular design which is scalable and robust for future implementations. The pipeline is not limited to a specific instrument although its suitability for data sets obtained in the mid-infrared (MIR) (3--5~$\mu$m) was a main requirement during its development. Background subtraction is critical in this wavelength regime and typically thousands of images have to be processed, making such data sets computationally more expensive to process compared to the optical and NIR regimes. Descriptions and results presented in this paper are based on PynPoint version 0.5.2.

The paper is structured as follows. In Sect.~\ref{sec:architecture}, we describe the new pipeline architecture, including the abstract interfaces and core functionalities. In Sect.~\ref{sec:pipeline_modules}, we outline some of the functionalities of the pipeline modules that are currently implemented to process and analyze data. In Sect.~\ref{sec:beta_pic}, we provide end-to-end examples of VLT/NACO coronagraphic data in $L'$ and VLT/NACO dithering data in $M'$ of the directly imaged planet $\beta$~Pictoris~b; we include a photometric and astrometric analysis and a quantification of detection limits. Finally, we summarize our work in Sect.~\ref{sec:summary}.

\section{PynPoint architecture}\label{sec:architecture}

The architecture of PynPoint has a modular design which separates the common data-handling functionalities that are required for all reduction steps from the actual data processing (see Fig.~\ref{fig:pipeline}). A simple pipeline interface is used to stack and run a sequence of various data-reduction algorithms given their input parameters. All data-reduction steps are capsuled with predefined inputs and outputs which sequentially fit together. The computational resources and the data sets themselves are managed by the back end of the package which provides the core functionalities related to storing and organizing data. It is possible to set the number of processes that run in parallel, as well as the number of frames that are simultaneously read into the memory. Therefore, PynPoint takes advantage of the available computational resources independent of the machine on which it is running (see Sect.~\ref{sec:hardware}).

\begin{figure*}
\centering
\includegraphics[width=0.75\textwidth]{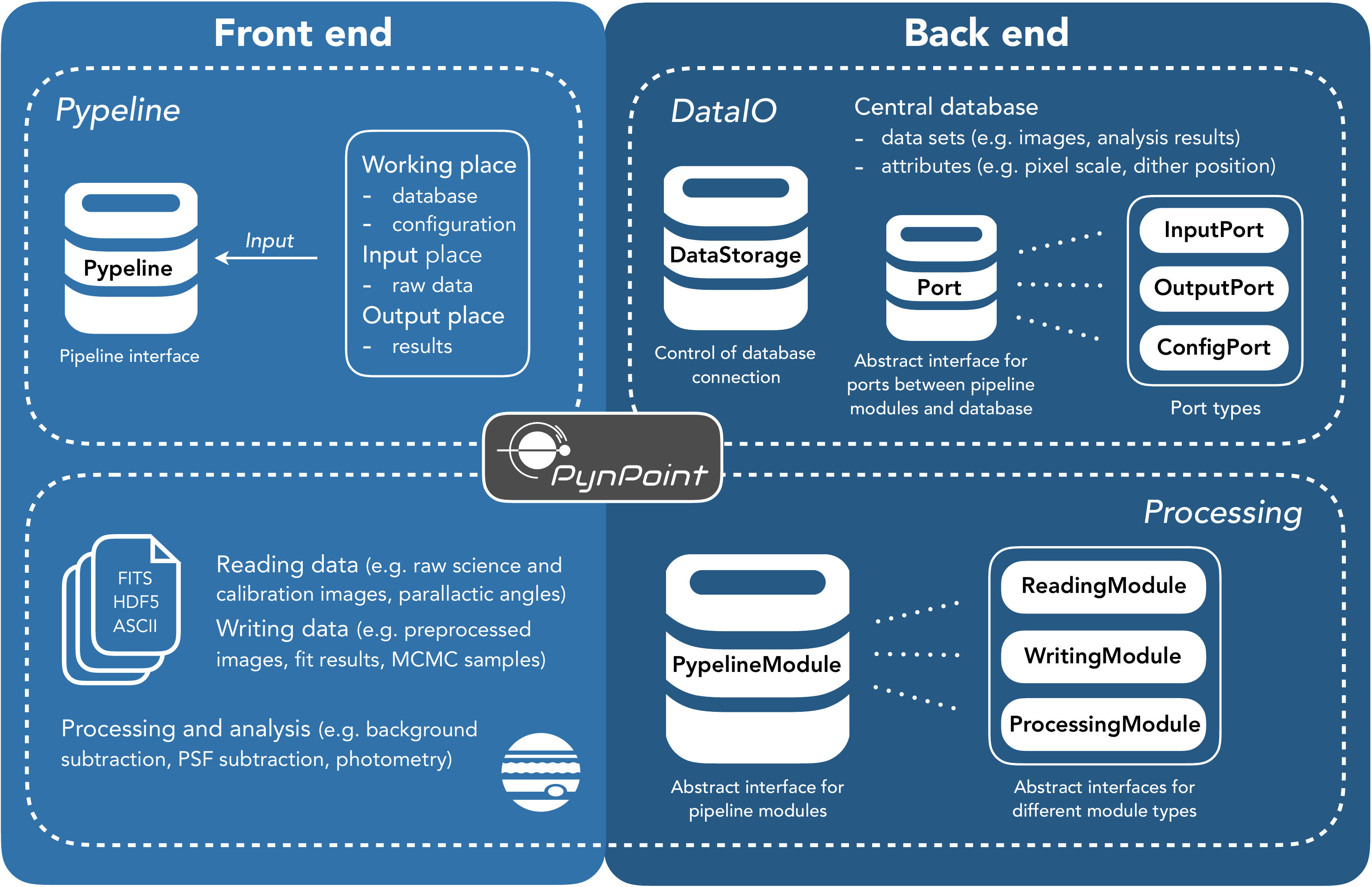}
\caption{Architecture of the core elements in PynPoint. The front end of the package contains the pipeline interface (\texttt{Pypeline}) which runs the pipeline modules that can read, process, and write data. The back end (\texttt{DataIO} and \texttt{Processing}) handles the connection to the central database (\texttt{DataStorage}) and provides the abstract interfaces for the ports and pipeline modules. There are interfaces for reading, writing, and processing modules, which all inherit from \texttt{PypelineModule}. Access to and storage of data occurs with the confined functionalities of the input, output, and configuration ports, which all inherit from the \texttt{Port} interface.\label{fig:architecture}}
\end{figure*}

The strict separation of data management and pipeline functionalities has several advantages. Firstly, the implementation overhead that is required to use the package for different data sets is small since the data is handled by the back end of PynPoint and the user can focus on the actual analysis. Secondly, the design makes the pipeline scalable to different data formats and new post-processing techniques while stability is ensured as new implementations will not intervene with the core functionalities. Finally, each pipeline module adds history information to the data sets that are processed by the pipeline. The complete history is stored as header information when a data set is exported, for example as a Flexible Image Transport System (FITS) file. This approach guarantees reproducibility of the results even if different pipeline modules and parameter values have been tested.

The front end of the software consists of the actual pipeline while the back end contains the interfaces for data input and output, and data processing. These functionalities are implemented in the \texttt{Pypeline}, \texttt{DataIO}, and \texttt{Processing} Python modules, respectively. A schematic overview of these core elements is shown in Fig.~\ref{fig:architecture} and is described in more detail below. Unit tests and style checks run automatically in the Github repository which ensures a robust implementation of new features while the execution of the core functionalities and existing pipeline modules remains unchanged.

\subsection{Pypeline -- the pipeline interface}\label{sec:pypeline}

The main interface of PynPoint is the \texttt{Pypeline} class which manages the data-reduction and analysis steps that have to be executed by the pipeline modules. The process starts with creating an instance of \texttt{Pypeline} which requires paths pointing to a working place, input place, and output place. The working place contains the central database in which all intermediate and final processing results are stored, as well as the configuration file which contains the central settings used by the pipeline and the relevant FITS header keywords that have to be read from the raw data. Each data set has a unique tag name used by the pipeline modules to select the requested data. The central database offers the flexibility to rerun certain processing steps by simply selecting the relevant tags of the data sets. The input place is the default location from where data are read into the central database and the output place is the default location where data and analysis results are written from the database (see Sects.~\ref{sec:data_io} and \ref{sec:io_modules}).

A \texttt{Pypeline} contains an internal dictionary of the pipeline modules with their unique name tags. Modules will be executed sequentially in the order by which they are added to the pipeline. It is possible to run all attached pipeline modules at once or a single module as specified by its name tag. A central database is created with the initialization of a \texttt{Pypeline} unless a database already exists in the working place. The configuration file contains a list of keywords and values which are read with the initialization of the pipeline and stored in a separate group in the database. A configuration file with default values is created if the file does not exist in the working place.

\subsection{Data input and output}\label{sec:data_io}

The \texttt{DataIO} module contains the classes that regulate (in the back end of the pipeline) the access of the modules to the central database. The \texttt{DataStorage} class is able to open and close the connection to the database in which the results from all the processing modules are written. The database is stored in the Hierarchical Data Format (HDF5) format which confers the advantage that reading slices of data and appending new data is much faster compared to the FITS format. In order to facilitate easy reruns of individual reduction steps and to simplify reproducibility, PynPoint stores the processing results from all pipeline modules. Therefore, sufficient disk space is required in the working place, in particular when the raw data contains thousands of images as is typical in the 3--5~$\mu$m range. For example, an HDF5 or FITS file with $10^{4}$ images of $1024 \times 1024$ pixels would require $\sim$79 GB of disk space.

Two types of data are stored in the database: data sets and attributes. Data sets are the main type of data that are processed, which typically contain a stack of images, but also analysis results such as fitted parameter values, detection limits, and samples from the Markov chain Monte Carlo (MCMC; see Sect.~\ref{sec:photastro}) are stored as data sets. Attributes are attached to a data set and contain header information. Some attributes are initially imported from the raw FITS data while others are automatically created with certain pipeline modules. Importing of attributes from the raw data is achieved with the keyword values provided in the configuration file. This is particularly useful if the user wants to work with data from different instruments, which only requires modifications in the configuration file. For example, setting the keyword \texttt{PARANG\_START} to the value \texttt{ESO ADA POSANG} implies that the values of \texttt{ESO ADA POSANG} in the FITS header are stored in the database as the nonstatic \texttt{PARANG\_START} attribute together the images. Alternatively, the attribute value of the pixel scale, \texttt{PIXSCALE}, can be set directly within the configuration file.

A distinction is made between static and nonstatic attributes. Static attributes are parameter values that are fixed for a data set (e.g., pixel scale, exposure time, and observatory location), as well as history information about the pipeline modules that were executed. Static attributes are attached to the data set itself and can change between data sets. For example, the pixel scale is adjusted when a resampling is applied to the images. Nonstatic attributes on the other hand are small data sets by themselves and contain arrays of parameter values that change between the imported data cubes (e.g., dither position, number of images in a FITS file, and exposure number) or on a frame-to-frame basis (e.g., parallactic angle, position of the star, and frame index). These types of attributes are stored in a separate group of the database and linked to an input port by the tag that is used for the corresponding data set. In this way, pipeline modules will automatically select the attributes belonging to the data set that is chosen as input data. All attributes are copied and updated, if needed, each time a pipeline module is executed.

Access to data in the central database is controlled by instances of \texttt{InputPort}, \texttt{OutputPort}, and \texttt{ConfigPort} which inherit the common port functionalities from an abstract interface called \texttt{Port}. Each port has an internal tag which works as a key to the central database and guarantees access to and changes of only the specified data set. The three different types of ports are implemented with specific and confined functionalities, which we summarize below.

An \texttt{InputPort} gives read-only access to a data set (typically a stack of images) in the central database. An input port also has the permission to read attributes that are associated with the data set. While an \texttt{InputPort} can only read data and attributes from the central database, an \texttt{OutputPort} can only store data in and delete data from the database. An output port sets up a connection to the database and can create a new data set or append data to an existing data set. Similarly, both static and nonstatic attributes can be added or overwritten, and nonstatic attributes can be appended to an existing list of values. Other functionalities of the output ports include comparing the name and values of static and nonstatic attributes, adding history information of a pipeline module, and copying all attributes from an input port. Finally, a \texttt{ConfigPort} reads the configuration values from the central database, for example the \texttt{PIXSCALE} (i.e., pixel scale of the image) or \texttt{MEMORY} (i.e., number of images that are simultaneously loaded into the memory).

\subsection{Data processing}\label{sec:processing}

Reading, writing, and processing of data occurs with the dedicated pipeline modules. As explained in Sect.~\ref{sec:pypeline}, these module are added to a \texttt{Pypeline} and executed sequentially. The implementation of modules occurs with the use of abstract interfaces of which \texttt{PypelineModule} is the overarching interface from which all types of modules inherit. This interface ensures that each pipeline module has a unique name tag as identifier, contains the obligatory functions to connect the ports of the pipeline module to the central database, and embeds the obligatory method to run the actual algorithm of the module.

There are three different types of pipeline modules: reading modules, writing modules, and processing modules. Each of them has its own abstract interface (\texttt{ReadingModule}, \texttt{WritingModule}, and \texttt{ProcessingModule}) which sets its functionalities, permitted ports to the central database, and obligatory parameters and methods. A reading module is only allowed to create output ports to the central database and is therefore suitable to read data from the hard drive and store them in the database. Data formats that are currently handled by the pipeline are FITS, HDF5, and ASCII tables (only for 1D and 2D data). Similarly, a writing module is only allowed to create input ports to the central database and can therefore be used to select data by their name tag and store them separately on the hard drive as a FITS, HDF5, or ASCII file. For example, a stack of preprocessed images can be stored as a FITS file or the parallactic angles as a ASCII table.

Processing and analysis of the data occurs with the processing modules of PynPoint. This type of module can setup one or multiple input and output ports to the database which enables them to access and store data sets. Each processing module has a dedicated task such as a flat-field calibration, background subtraction, bad-pixel correction, PSF subtraction, or flux and position measurement. With the design of the abstract interfaces for the various pipeline modules, one can easily implement new pipeline modules with the available functionalities to create input and output ports from a module to the central database, to execute the actual processing algorithm with the run method, and to update attributes in the database.

\subsection{Management of hardware resources}\label{sec:hardware}

A central component of PynPoint's back end is the management of available hardware resources (i.e., memory and processors). It is possible to read, process, and write the images in subsets instead of loading all the images from a data set into the memory at once. The number of images per subset is provided in the configuration file with the \texttt{MEMORY} keyword. This approach enables the processing of thousands of images, as is typical in the 3--5~$\mu$m range, without overloading the computer memory, and confers the advantage that data can be decomposed and processed in parallel.

\begin{figure}
\centering
\includegraphics[width=\linewidth]{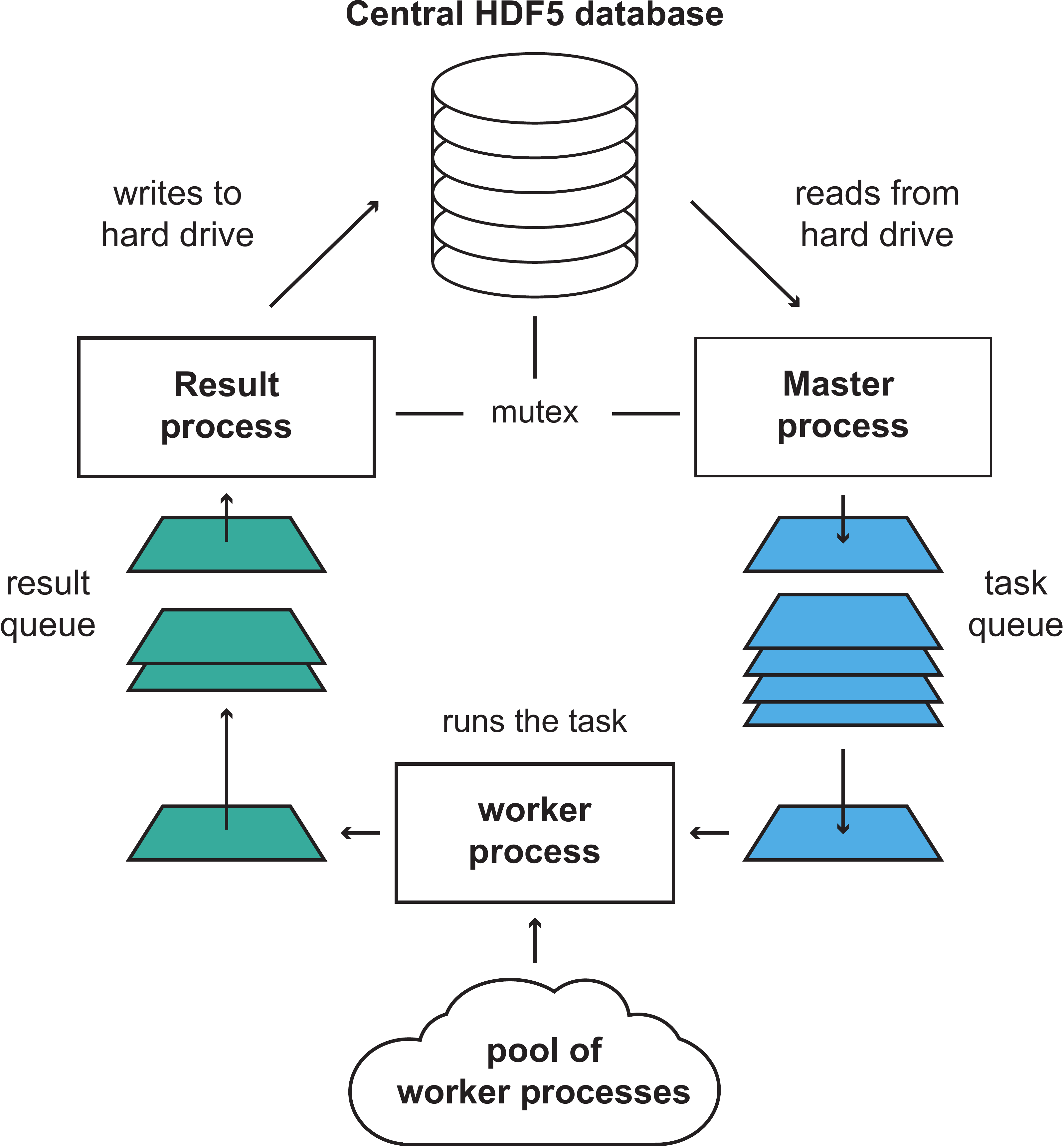}
\caption{Multiprocessing scheme of PynPoint. The workflow consists of two processes for read and write operations and a pool of worker processes to run the actual algorithm in parallel. In this manner, not only is the computation time reduced but the time required for reading and writing of data can also be hidden during the computations. See main text for details on this so-called master-worker pattern.\label{fig:multiprocessing}}
\end{figure}

Parallel processing of data sets in PynPoint follows the master-worker pattern \citep{mattson2004} which is illustrated in Fig.~\ref{fig:multiprocessing}. This approach allows for a flexible number of parallel processes while limited latency occurs due to read  and write operations. A so-called master process starts to read subsets of the complete input data set from the hard drive and transfers them into a fixed-size task queue in the memory. A pool of worker processes waits for data to be present in the task queue in order to execute the desired algorithm in parallel. Afterwards, each process transfers the result back into a result queue and waits for a following subset to be processed. Finally, a writer process waits for results to appear in the result queue and writes them back to the hard drive. The number of processes that will run in parallel can be specified by the \texttt{CPU} keyword in the configuration file.

To prevent simultaneous read and write access to the central database, a mutual exclusion (mutex) variable is used by the processes. Specifically, the master process is blocked if the writer process is storing the result in the database and the writer process is blocked if the master process is reading from the database. After the complete data set has been read by the master process, it places so-called poison pills into the task queue which will shut down the workers. The worker processes operate independently of the read and write operations, therefore the latency caused by the input and output processes can be hidden. Latency hiding means that the time required to read and write data from and to the hard drive does not have a significant impact on the overall run time as these operations are executed in parallel to the processing of the data. Overhead only occurs at the start and end as there will be some time during which only read and write processes are active.

Processing modules often apply a specific procedure on all images of a data set. In order to enable future implementations to benefit from the memory management and parallel implementations, a generic method is available to process all images of a data set with a specified function, which can be called from any pipeline module. Similarly, there are methods available to perform, for instance, multiple PCA fits at the same time and for processing multiple pixels in the time domain as is used by the wavelet-based speckle suppression \citep{bonse2018}.

\section{Pipeline modules}\label{sec:pipeline_modules}

A range of processing modules are currently implemented for various data-reduction steps. Some of the modules were specifically developed for data obtained in the \mbox{3--5~$\mu$m} range but most modules are also suitable for data sets obtained at optical or NIR wavelengths. In this section we briefly summarize some of the main features of the processing modules but we refer to the online documentation\footnote{\url{http://pynpoint.readthedocs.io}} for a more complete overview of the processing functionalities and a description of the input parameters. As PynPoint is under continuous development, alternative algorithms for the various data-reduction steps could be considered.

\subsection{Importing and exporting data}\label{sec:io_modules}

Several reading and writing modules are available to import data in and export data from the central database. The supported data formats are FITS, HDF5, and ASCII tables but the abstract interfaces allow for an easy implementation of input and output modules that support other data formats. Since the central database is stored in the HDF5 format, it is possible to export one or multiple tags from the database to a separate HDF5 file with an automatic inclusion of the associated attributes. This is in particular useful if the end product of a processed data set is used as input product for a different pipeline (e.g., PSF template for the computation of detection limits). Similarly, one or multiple data tags from an external HDF5 file can be imported, together with the attributes, into the central database.

Images from FITS files are read together with the relevant header keywords that are required for processing of the data. The central configuration file is used to link keywords in the FITS header to the associated attributes as used by PynPoint. For example, the nonstatic attribute that contains the exposure number is \texttt{EXP\_NO}, used for the background subtraction of data obtained with nodding (e.g., VLT/NACO annular groove phase mask (AGPM) data; see Sect.~\ref{sec:naco_lp}), which in the FITS header of data obtained with VLT/NACO is given by \texttt{ESO DET EXP NO}. In this way, both the images and required header keywords are read from the FITS files and stored as data sets and attributes in the central database. Reading of FITS header keywords is optional as whether or not  certain keyword values are required for the data reduction depends on the data. Similarly, data sets from the database can also be exported to FITS files such as a stack of processed images that is required for further processing with other tools.

There are also reading and writing modules available for ASCII tables which is currently only supported for 1D and 2D data. For example, a list of parallactic angles can be read and attached as the \texttt{PARANG} attribute to a data set or detection limits can be exported from the database to a separate ASCII file.

\subsection{Preprocessing and cosmetics}\label{sec:preprocessing}

Basic calibration modules are available for dark-current subtraction and flat-field correction. For a first-order distortion correction it is possible to scale images in both image dimensions while conserving the total flux. In the 3--5~$\mu$m range, subtraction of the thermal background emission from the sky, telescope, and instrument is important, so several modules are dedicated to this procedure. For data obtained with dithering, the background emission can be subtracted with a mean background frame that is created from the previous and/or subsequent data cube in which the star has shifted to a different position. The background frames of the adjacent data cubes are selected with the \texttt{NFRAMES} attribute which contains the number of frames in each imported FITS file. The \texttt{NFRAMES} attribute is updated whenever frames are removed from a data set.

Alternatively, a background subtraction algorithm based on PCA can be applied as introduced by \citet{hunziker2018}. This method may provide particularly good results with varying observing conditions and/or infrequent sampling of the sky background. In case of dithering data, the dither positions can be selected and sorted as \emph{star} and \emph{background} frames with the \texttt{DITHER\_X} and \texttt{DITHER\_Y} attributes. A mean background subtraction (based on the adjacent data cubes) is temporarily applied to locate the star more easily such that the PSF can be masked. Next, the background frames are decomposed into an orthogonal basis and the frames containing the star are fitted both in the masked and nonmasked region. Optionally, the average frame of the entire stack with background frames is subtracted from both the star and background frames before the PCA basis is created and fitted. More details and examples about the PCA-based background subtraction scheme are provided in \citet{hunziker2018}.

A separate background subtraction module is implemented for data obtained with nodding (e.g., VLT/NACO AGPM data). For such data sets, the sky background is sampled by alternating the pointing of the telescope between the science target and an empty region on the sky. Each cube of sky frames is averaged and subsequently subtracted from the science frames based on the \texttt{EXP\_NO} attribute. This pipeline attribute can be assigned to a keyword in the FITS header that identifies the order in which the data cubes were written. Background subtraction occurs by selecting the subsequent, previous, or average of both sky frames that are nearest in time to the science frames. Alternatively, it is also possible to apply the PCA-based background subtraction by creating the PCA basis from the sky frames and fitting each science frame with a linear combination of the basis components.

Bad pixels can be corrected with two different methods. The fast approach is by sigma clipping of pixel values deviating more than a specified number of standard deviations within a given filter size and replacing them by the mean of the neighboring pixel values. This approach may work well for single outliers but will not correct clusters of bad pixels.

In order to correct both single outliers and clusters of bad pixels we have adopted a spectral deconvolution algorithm, which was first presented by \citep{franke1987}. The algorithm replaces bad pixels in an iterative manner with the dominant frequencies from the Fourier representation of the individual frames. Spectral deconvolution takes advantage of the fact that a point-wise multiplication in image space is equal to a convolution in frequency space. Local defects in image space are spread in frequency space and have a smaller impact on the individual frequency values. Hence, the individual frequencies are only slightly affected by bad pixels and can be used for interpolation. The module requires a bad-pixel map as input, which can, for instance, be constructed (with a separate module) from the dark- and flat-field frames. Alternatively, the module for sigma clipping produces a bad-pixel map which may also capture nonstatic bad pixels.

The spectral deconvolution algorithm starts by selecting dominant frequencies in the Fourier representation of an image and transforms them back to the image space. The result is a first approximation of the interpolated pixel values that will replace the flagged pixels in the bad-pixel map. After an update of the Fourier representation the defects in frequency space will be less dominant. The iteration continues with the updated representation and creates an updated approximation of the pixel values by selecting additional dominant frequencies, thereby increasing the interpolation accuracy of the defect pixel regions. To speed up the computation and to improve the numerical stability, we have adopted, from \citet{aach2001}, a slight variation of the original algorithm.

\begin{figure}
\centering
\includegraphics[width=\linewidth]{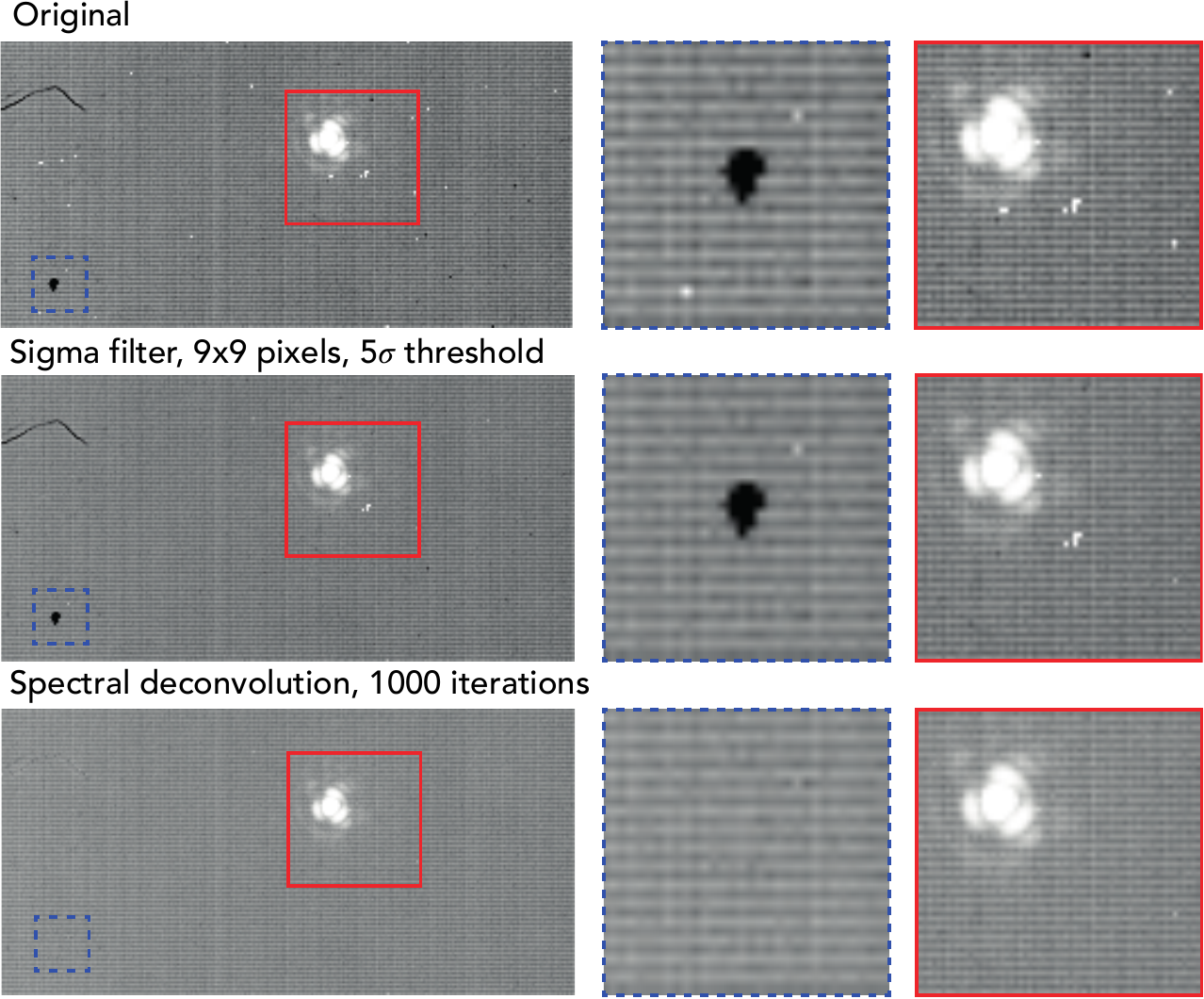}
\caption{\emph{Top:} Raw image of a 0.2~s exposure of $\beta$~Pictoris obtained with VLT/NACO in the $L'$ filter. The close-up of the blue region shows a cluster of bad pixels and the close-up of the red region shows the PSF of the star with several surrounding bad pixels. \emph{Center:} Sigma clipping applied to the raw image with a filter of $9 \times 9$~pixels and a threshold of $5\sigma$. \emph{Bottom:} Interpolation of the preselected bad pixels with 1000 iterations by the spectral deconvolution.\label{fig:bad_pixels}}
\end{figure}

A comparison of the two methods for bad-pixel corrections is displayed in Fig.~\ref{fig:bad_pixels}, showing a raw frame of archival VLT/NACO data in the $L'$ filter (ESO program ID: 084.C-0739(A)). While single pixel outliers are well corrected with the sigma filter, small clusters of bad pixels are not. By first constructing a bad pixel map from the dark- and flat-field images and subsequently applying the spectral deconvolution algorithm with 1000 iterations, we are able to correct both single outliers and clusters of bad pixels. The corrected image shows a continuous detector pattern with no obvious artifacts introduced by the interpolation of the bad pixels.

In addition to correcting bad pixels, a frame selection of the data might be required in case part of the data was obtained under variable conditions during which the AO performance plummeted or the AO loop opened. This is implemented by measuring the total flux in a circular or annular aperture (or the ratio of the two, that is, the halo-to-core flux ratio of the PSF) either at a fixed position or by first locating the position of the star. Subsequently, frames are removed that are more than a given number of standard deviations away from the maximum or median photometry value.

\subsection{Image registration}\label{sec:image_registration}

After basic calibration and cosmetic corrections, images have to be registered by placing the stellar PSF in the center of the image. For noncoronagraphic observations, the star can be located by selecting the highest pixel value (in a subsection of the image if needed), after smoothing the image with a Gaussian filter similar in size to the stellar PSF to lower the contribution of possibly remaining hot pixels. This step also crops the images to a specified size which will speed up the computation time for the remaining processing steps. Alignment of the images is done by cross-correlating each image with a specified number of random images from the same stack or a different data set in the database. Optionally, an upsampling can be applied before shifting the images to the average offset value with a default fifth-order spline interpolation.

Alternatively, images can be centered by fitting a 2D Gaussian profile in cases where the core of the PSF is unsaturated. A circular region can be specified around the approximate PSF center which is used for the nonlinear least-squares fit with a Levenberg-Marquardt algorithm \citep{levenberg1944,marquardt1963}. After the fit, the image is shifted and the procedure is repeated for all other images in the stack. When images have already been aligned in a previous step with a cross-correlation, a fit to the mean of the image stack can be applied and all images are shifted by a constant offset in order to have them centered with subpixel precision. A coronagraph is commonly used for pupil-stabilized observations in which case a different centering approach has to be applied. For example, the PSF of data obtained with an AGPM coronagraph \citep{mawet2005} is annular in its morphology as a result of the vortex mask. The central hole of the annulus can be fitted with a 2D Gaussian by changing the sign of the pixel values and shifting all values by a constant offset to positive values. Images are then shifted with subpixel precision by assuming that the star is located at the center of the annular residuals.

\subsection{Point spread function and speckle suppression}\label{sec:psf_subtraction}

PynPoint was originally developed as a PSF-subtraction tool with PCA \citep{amara2012,amara2015}. Full-frame PCA remains the main and currently only implementation for PSF subtraction but the modular and scalable architecture allows for easy implementation of different PSF-subtraction modules which may provide higher sensitivity compared to PCA. As advances are made to improve post-processing algorithms for pupil-stabilized data \citep[e.g.,][]{gomez-gonzalez2017}, we designed the pipeline such that the architecture is suitable for additional PSF subtraction techniques. A separate module is available to prepare the PSF subtraction by masking the inner and/or outer regions of an image and optionally normalizing each image by its Frobenius norm. Furthermore, it is possible to stack and/or randomly sample the images prior to the PSF subtraction with a dedicated pipeline module.

The PSF subtraction module requires that the \texttt{PARANG} (parallactic angle) attribute is attached to the data set, which can be achieved in different ways. For example, the parallactic angles can be imported from an ASCII table (see Sect.~\ref{sec:io_modules}), estimated with a linear interpolation between the start and end value of each data cube, or calculated more precisely from the relevant header information. The latter two approaches require input in the configuration file; for example, the interpolation uses the values from the \texttt{PARANG\_START} and \texttt{PARANG\_END} attributes which should be linked to the relevant FITS header keyword before the raw data are imported (see Sect.~\ref{sec:data_io}).

The PSF subtraction with PCA is implemented with the functionalities of \texttt{scikit-learn} \citep{scikit-learn2013}, a Python package for machine learning, from which we have chosen the ARPACK library for the singular value decomposition (SVD). A range of principal components (PCs) can be specified and the orthogonal basis is constructed by decomposing the stack of images onto the lower dimensional space of the maximum number of specified PCs. The best-fit linear combination of the PCs is then calculated for each image and the PSF model is subtracted from the data. Images are then derotated to a common field orientation with the \texttt{PARANG} attribute, mean and/or median combined, and written to the specified database tag.

Creating, derotating, and stacking can be computationally expensive if the stack contains thousands of images and the residuals have to be obtained for a range of PCs. Therefore, this processing step can be executed in parallel with the multiprocessing implementation (see Sect.~\ref{sec:hardware}). This is only possible using a machine with sufficient memory because each process requires enough memory to store a copy of the PCA basis.

A new speckle-suppression technique was recently introduced by \citet{bonse2018} which is based on wavelet transformations. This technique uses the frequencies and time dependence of the speckle variations to filter out the speckle noise. Wavelet denoising is not a replacement for the PSF-subtraction module but can be applied as an additional preprocessing step earlier on in the data-reduction sequence. Improvements in sensitivity are in particular expected if the temporal sampling is large (i.e., short exposure times) and the data cover a large variation in parallactic angle. \citet{bonse2018} demonstrated that in such cases improvements of the signal-to-noise ratio (S/N) can be as large as 40--60\%.

\subsection{Analysis tools}\label{sec:analysis_tools}

In addition to pipeline modules for pre- and post-processing, there are several modules implemented for analyzing reduced data, including modules for injection of artificial point sources (referred to as planets from here on), estimation of detection limits, and photometric and astrometric analysis of a point source.

\subsubsection{Artificial planets and detection limits}\label{sec:detection_limits}

Artificial planets are injected by providing a database tag that points to the centered science data and a PSF template. The tag of the science data and PSF template can be identical, for example when the data were obtained without coronagraph and remained unsaturated throughout the sequence. In that case an exact copy of each science image can be used, which will yield the most precise realization of an artificial planet. However, this is often not possible for a given observing strategy such that unsaturated images are separately obtained, for example before and after a coronagraphic sequence. In that case an identical PSF template is injected in each science image, which therefore reduces the photometric accuracy since changes in observing conditions cannot be accounted for.

The PSF template is shifted with a fifth-order spline interpolation in horizontal and vertical directions to the specified separation and position angle while taking into account the parallactic rotation. The flux is scaled by the specified contrast with the star and an additional scaling can be applied, for example to correct for a difference in detector integration time (DIT) or the transmission of a neutral density (ND) filter. Both positive and negative planet signals can be injected which are required for estimating detection limits and photometric measurements of a real planet signal, respectively.

A contrast curve provides detection limits at a fixed $\sigma$ threshold as a function of separation from the star. This means that the false positive fraction (FPF) decreases (and the related confidence level increases) with increasing separation as the number of independent reference apertures (with a diameter of $\geq\lambda/D$) increases towards larger separations, following the small sample statistics of the Student's t-distribution \citep{mawet2014}. Therefore, $\sigma$ is only associated with a $\sim$68\% confidence level at large separations where the sampled distribution approaches Gaussian statistics. We note that the contrast curve is not a robust metric for presenting and comparing detection limits, so alternative implementations could be considered such as the recently proposed performance map \citep{jensen-clem2018} or a derivation of upper limits in a Bayesian framework \citep{ruffio2018}. We leave this for future work and for now adopt the traditional approach of the contrast curve which applies a correction for small sample statistics but fixes the $\sigma$ level with separation.

The detection limits are estimated through an iterative process of injecting an artificial planet, subtracting the stellar PSF with PCA, calculating the FPF, and comparing the value with the FPF associated with the specified $\sigma$ threshold (at that separation). Multiple iterations of the PSF subtraction are required at each position until the FPF has converged to a specified accuracy. The process is repeated at a range of separations and position angles which are specified by a minimum and maximum value, and a step size. Although the iterative process makes it computationally expensive, we chose this approach because the algorithm throughput may depend on the brightness of the artificial planet. Calculating the throughput for a single brightness and scaling the value to the limiting contrast may therefore bias the result. The number of PCs is fixed with separation, which means that self-subtraction effects become more severe at smaller separations because the artificial planet rotates along a shorter path and the required brightness increases due to enhanced speckle noise at smaller separations. Such effects can be investigated by calculation of detection limits for a range of PCs. The results that are stored in the central database include the azimuthally averaged detection limits, the azimuthal variance of the limits, and the FPF.

\subsubsection{Photometry and astrometry of companions}\label{sec:photastro}

Photometric and astrometric measurements of directly imaged companions are challenging due to self-subtraction effects that are inherent to the current approaches to analyze ADI data sets. Negative artificial planets are commonly used to determine the planet's position and brightness relative to its star \citep[e.g.,][]{marois2010b,lagrange2010}. In PynPoint, this is achieved by subtracting a scaled replica of the stellar PSF at the location of the planet and iteratively minimizing the residuals with a downhill simplex method \citep{nelder1965}. Two different merit functions are currently implemented for the simplex minimization.

The first merit function considers the curvature of the image surface, which can be quantified by the determinant of the second-order derivatives, and is defined as
\begin{equation}\label{eq:hessian}
f = \sum_{i,j}^N \left|\det{\left(\mathbf{H}_{ij}\right)}\right|,
\end{equation}
where $i$ and $j$ are the pixel indices, $N$ is the total number of pixels encircled by a circular aperture at the fixed, approximate position of the planet, and $\mathbf{H}_{ij}$ is the Hessian ($2 \times 2$) matrix which gives the second-order partial derivatives at each pixel position. The determinant of the Hessian matrix is referred to as the Hessian and is used to analyze critical points in a matrix, and is therefore a measure for the curvature of the image surface. Taking the absolute value of the Hessian ensures that both local minima and maxima give a positive value of the Hessian whereas a flat surface (i.e., in case of a perfect subtraction) reduces the Hessian towards zero. Optionally, the residuals of the PSF subtraction can be convolved with a Gaussian kernel prior to the computation of the Hessian in order to reduce pixel-to-pixel variations. The impact of the width of the Gaussian on the photometric and astrometric precision is investigated in Sect.~\ref{sec:beta_pic_local}.

The second merit function considers the flux values of the image residuals and is defined by a $\chi^2$ function \citep{wertz2017,gomez-gonzalez2017},
\begin{equation}\label{eq:sum}
f = \sum_{i,j}^N \left|I_{ij}\right|,
\end{equation}
where $i$ and $j$ are the pixel indices, $N$ is the total number of pixels encircled by the aperture, and $I_{ij}$ is the pixel value with the exact fractional overlap of pixels taken into account. Here it is assumed that the image residuals after PSF subtraction are equal to zero and that the uncertainties on the pixel counts follow a Poisson distribution such that $\sigma_{ij} = \sqrt{I_{ij}}$. Equation~\ref{eq:sum} assumes uncorrelated measurement uncertainties which is not strictly true for neighboring pixel values because quasi-static speckle noise evolves on various timescales \citep[e.g.,][]{macintosh2005,hinkley2007}.

The minimization scheme adjusts the position and brightness of the negative artificial planet (after an initial guess) and each step runs the PSF subtraction and computes the merit function within a circular aperture. This procedure is continued until the absolute error is smaller than the specified acceptance threshold. The image residuals of each iteration are stored in the central database to make them available for visual inspection. In Sect.~\ref{sec:beta_pic_local}, we provide an example analysis of the photometric and astrometric precision that is achieved with minimization of the determinants of the Hessian.

The simplex minimization provides a fast way for determining the relative brightness and position of a detected point source. To explore the parameter correlations and estimate uncertainties on the separation, position angle, and contrast, a pipeline module is available for an MCMC analysis with the affine-invariant ensemble sampler implementation of the \texttt{emcee} package \citep{foreman2013}, as proposed by \citet{goodman2010}. The results from the simplex minimization can be used as a starting point for the walkers. The position of each walker in the ($\rho$, $\theta$, $\delta$) space is initialized by a random value from a Gaussian distribution. For each step the walkers sample a new value of the angular separation, $\rho$, the position angle measured counterclockwise with respect to the upward direction, $\theta$, and the planet-to-star flux contrast, $\delta$.

The posterior probability function is proportional to the product of the prior probability function and the likelihood function. Here we impose uniform priors on the three parameters and transform Eq.~\ref{eq:sum} into a log likelihood function, $\ln \mathcal{L} = -\frac{1}{2} \chi^2$, following \citet{wertz2017}. Convergence of the chains is tested with the integrated autocorrelation time, $\tau_\mathrm{int}$, which measures the correlation of the chains at different steps of the walkers. Therefore, a low autocorrelation indicates that the parameter space is efficiently sampled such that the posterior distribution is close to the ground truth. The pipeline module runs the MCMC simulation in parallel through the specification of the number of processes in the configuration file. The gain in computation time depends on the expense of the log likelihood function and the size of the stack of images specifically.

\section{Application to archival data of $\beta$~Pictoris}\label{sec:beta_pic}

In this section we describe two end-to-end examples and present the analysis results of the processed data. The examples are based on archival data of $\beta$~Pictoris, an A6V type star which is orbited by the gas giant planet $\beta$~Pic~b \citep{lagrange2009,lagrange2010}. The data were obtained with VLT/NACO in the $L'$ and $M'$ filters, therefore requiring special attention to be paid to the background subtraction. All the processing steps that are described in Sects.~\ref{sec:naco_lp} and \ref{sec:naco_mp}, as well as the analysis presented in Sects.~\ref{sec:beta_pic_photometry} and \ref{sec:beta_pic_limits}, are carried out with pipeline modules that are implemented in \mbox{PynPoint}.

\subsection{VLT/NACO $L'$ coronagraphic data}\label{sec:naco_lp}

The first data set was obtained with VLT/NACO in the $L'$ filter ($\lambda_0 = 3.80$~$\mu$m) on UT 2013 February 01 (ESO program ID: 60.A-9800(J)) and presented by \citet{absil2013}. The data were taken during science verification after the installation of the AGPM coronagraph. Observations were executed under mediocre conditions with frequent opening of the AO loop. The background was sampled by nodding the telescope to a nearby sky region during the science sequence.

The DIT was 0.2~s with 200 and 50 exposures per cube of science and sky data, respectively. Sequences of 10 or 20 cubes of science data were taken with three cubes of sky frames in between, resulting in a sampling of the sky background every $\sim$10~min. The combination of the DIT and the window size of $768 \times 770$~pixels caused a $\sim$10\% frame loss\footnote{See Section 5.7 in the VLT/NACO User Manual, Issue: 101} for each data cube that was written. Unsaturated exposures of the stellar PSF were obtained with a DIT of $\sim$0.02~s before and after the coronagraphic sequence by shifting the star away from the coronagraph. Fitting of a 2D Gaussian function to the stellar PSF yielded a mean full width at half maximum (FWHM) of 114~mas ($\lambda/D = 96$~mas). Data were obtained in pupil-stabilized mode to take advantage of ADI, with a total parallactic rotation of $83\ffdeg1$. We refer the reader to \citet{absil2013} for further details on the observations.

The data reduction started by importing the raw science (34277 images), sky (1989 images), flat-field (15 images), dark-current (3 images), and unsaturated exposures (4859 images) to separate tags in the central database. The unsaturated images were reduced separately but the processing steps, including a frame selection which removed 21\% of the images, were similar to the ones described both here and in the following section. The last frame from each imported science and sky data cube was removed because it contained the average of the individual exposures that were stored with the cube mode of NACO. The field orientation of the individual images was calculated with a linear interpolation between the start and end values of the position angle on sky for each data cube. A precise calculation of the parallactic angles was not possible because of the frame loss when the data were written from the detector.

\begin{figure*}
\centering
\includegraphics[width=\linewidth]{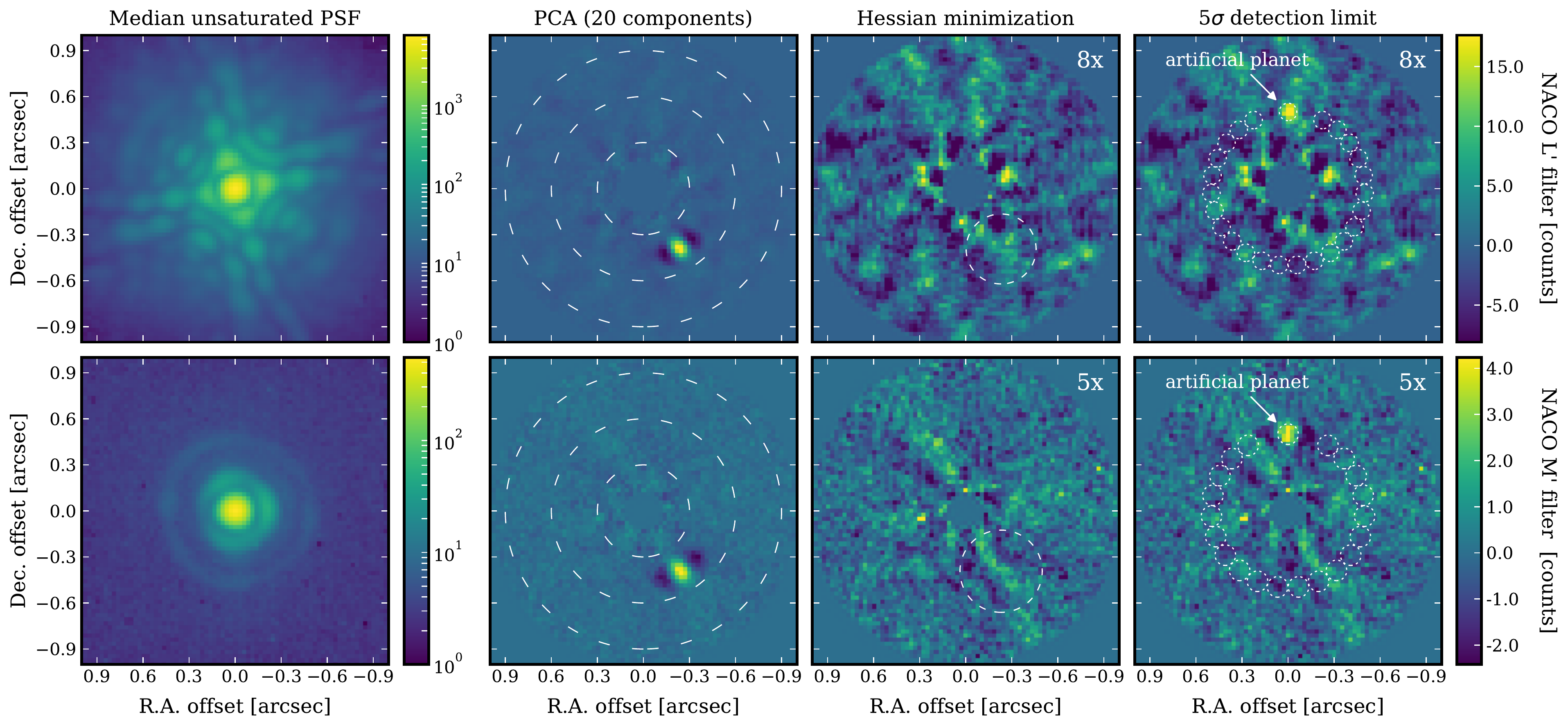}
\caption{Analysis of archival VLT/NACO data of $\beta$~Pictoris in the $L'$ (\emph{top}) and $M'$ (\emph{bottom}) filters. Images show from \emph{left} to \emph{right} the median unsaturated PSF of $\beta$~Pic, the median residuals of the PSF subtraction, the residuals with the best-fit negative artificial planet injected, and the residuals with an artificial planet injected at 0\ffarcs5 and a brightness scaled to a $5\sigma$ detection level (i.e., a false positive fraction of $3.97 \times 10^{-5}$ at that separation). All residuals were obtained by fitting the images with 20 PCs. The dashed circles in the second column are there to guide the eye, the dashed circles in third column denote the aperture size that was used for the minimization, and the dotted circles in the fourth column show the position and size of the detection and reference apertures. The reference apertures are used to estimate the noise at the separation of the artificial planet. The unsaturated images and the PSF subtraction residuals are displayed on a logarithmic and linear color scale, respectively. The unsaturated flux is shifted by a few counts such that the minimum value is unity. The dynamical range of the minimization and detection limits residuals is enhanced by the factor in the top right corner of each image, compared to the color bars on the right. North is up and east is to the left in all post-processed images.\label{fig:images}}
\end{figure*}

The top two pixel rows of the detector do not contain useful data so they were removed such that images are square. The science and sky background data were then divided by the master flat-field which was created from the sky flats with a subtraction of the dark-current and a normalization of the averaged images. The first five frames in each cube of science and sky data were removed (based on the \texttt{NFRAMES} attribute) because the background emission had a systematic offset at the start of all data cubes and decreased exponentially to a constant level during the sequence. Images were cropped to $5\arcsec \times 5\arcsec$ and the mean sky background was subtracted. This was achieved by first averaging the sky images of each sky data cube and then computing the mean of the previous and subsequent sky frame closest in time to each science image.

In addition, we also processed the images with a PCA-based background subtraction in order to compare the detection limits with both background subtraction schemes in Sect.~\ref{sec:beta_pic_limits}. The stack with the sky images was decomposed into a basis set of 60 PCs \citep[cf.][]{hunziker2018} which was then used to fit a linear combination of the PCs to the background of each science image. The inner 4 FWHM radius around the star was excluded for the construction of the PCA basis but included in the fit and background subtraction.

In the next step, we corrected bad pixels by three times iterating with a sigma filter of $9 \times 9$~pixels and a threshold of $5\sigma$. This procedure selected and corrected on average 11.4 pixels per image with a size of $185 \times 185$~pixels. Low-quality frames were removed by measuring the ratio of the integrated flux in an annular aperture, with an inner and outer radius of 1 and 4~FWHM, and a circular aperture with a radius of 1~FWHM, both centered on the star. Frames with flux ratios deviating by more than $1.5\sigma$ from the median flux ratio were removed, corresponding to 11\% of the frames. The threshold was iteratively chosen by visual inspection of the removed frames.

The star had been recentered behind the coronagraph mask each time the sequence changed from sky to science exposures which is routinely achieved with a precision of $0.1\lambda/D \simeq 10$~mas (VLT/NACO User Manual, Issue: 101). Therefore, we assumed that the science and sky frames (which include thermal emission from the coronagraph) were well aligned. Instead, only an absolute centering was done by averaging the stack of images, upsampling the averaged image by a factor of five, fitting a negative 2D Gaussian to the central region (1~FWHM in radius), and finally shifting all images by the computed offset in both image dimensions. The stack of images appeared well centered on visual inspection of the central annulus.

To speed up the MCMC sampling procedure and the calculation of the detection limits later on, we mean-combined every 50 images and their parallactic angles, resulting in a final stack of 594 preprocessed images. Pixel values within a radius of 1.5~FWHM and beyond $1\ffarcs0$ were masked before the stellar PSF and quasi-static speckles were modeled and subtracted with the PCA implementation of the PSF subtraction. Image residuals were then derotated to a common field orientation, and mean- and median-combined. The median image residuals obtained with 20 PCs are shown in the top row in Fig.~\ref{fig:images} (second image from left). The S/N was computed from the median-combined residuals for 1--50~PCs with an aperture of 1~FWHM in diameter centered on the approximate position of the planet, accounting for small sample statistics \citet{mawet2014}. This procedure yielded maximum S/Ns of $\sim$20 (26~PCs) and $\sim$23 (27~PCs) when either a mean or PCA-based background subtraction was applied, respectively.

\subsection{VLT/NACO $M'$ dithering data}\label{sec:naco_mp}

The second data set of $\beta$~Pictoris that was reprocessed and analyzed had been taken with VLT/NACO in the $M'$ filter ($\lambda_0 = 4.78$~$\mu$m) on UT 2012 November 26 (ESO program ID: 090.C-0653(D)) and was presented by \citet{bonnefoy2013}. A four-point dither pattern was applied during the observations to sample the sky background such that a different approach for the background subtraction is required compared to the coronagraphic $L'$ data.

We started by importing the raw science (55384 images), flat-field (5 images), dark-current (3 images), and unsaturated exposures (1960 images) to separate tags in the central database. The selected science data consisted of 184 data cubes with each cube containing 300 exposures of 65~ms. The unsaturated images of the stellar PSF were obtained with a similar dither pattern and processed separately from the science data. The mean FWHM of the unsaturated stellar PSF is 134~mas ($\lambda/D = 121$~mas) as determined with a 2D Gaussian fit. The observations were executed in pupil-stabilized mode with a total field rotation of $51\ffdeg6$.

Similar to the $L'$ data, every NDIT+1 frame was removed from each imported data cube because it contained the average of the cube. The parallactic angle, $\pi_i$, at observing time $t_i$, was calculated as
\begin{equation}\label{eq:parallactic}
\pi_i = \arctan{ \left( \frac{\sin{h_i}}{\cos{\delta}\tan{\phi}-\sin{\delta}\cos{h_i}} \right) },
\end{equation}
where $i$ is the index of the image, $h_i$ is the hour angle, $\delta$ is the declination of the target, and $\phi$ the geographical latitude of the observatory. The target coordinates, location of the observatory, DIT, and UT time at the start of a data cube were read from the FITS headers based on the specification in the configuration file. There are no additional overheads that have to be considered with NACO's cube mode. The calculated angles were corrected for the position angle of the telescope pupil and the constant rotator offset of $89\ffdeg44$\footnote{See Section 5.8 in the VLT/NACO User Manual, Issue: 101}. In the derotated images, north and east will be pointing in upward and leftward direction, respectively. Again, the top two pixel lines were removed, a flat-field correction was applied, and the first five frames were removed because of the systematically higher background.

The four dither positions, which typically correspond to the four detector quadrants, are then cropped from the full detector array to $3\ffarcs5 \times 3\ffarcs5$ and sorted as star or background frame. We then applied the PCA-based background subtraction by placing a mask (4~FWHM in radius) at the approximate position of the star and decomposing the stack of background images onto the first 60 PCs, for each dither position separately. The best-fit linear combination of the basis components was computed and subtracted for each image separately. Additionally, we processed the data with a mean background subtraction based on the adjacent data cube, as described in Sect.~\ref{sec:preprocessing}.

\begin{table*}
\caption{Photometric and astrometric analysis of $\beta$~Pic~b with PynPoint.}
\label{table:beta_pic_analysis}
\centering
\bgroup
\def\arraystretch{1.25}
\begin{tabular}{L{6cm} C{3cm} C{3cm} C{3cm}}
\hline\hline
Method & Contrast (mag) & Separation (mas) & Position angle (deg) \\
\hline
 & \multicolumn{3}{c}{VLT/NACO $L'$ filter} \\
Hessian minimization, 10 PC & 7.88 & 449.46 & 210.20 \\
Hessian minimization, 20 PC & 7.90 & 449.52 & 210.27 \\
Hessian minimization, 30 PC & 7.88 & 450.12 & 210.25 \\
Hessian minimization, 40 PC & 7.91 & 449.88 & 210.27 \\
Flux minimization, 10 PC & 7.86 & 450.33 & 210.18 \\
Flux minimization, 20 PC & 7.84 & 449.41 & 210.26 \\
Flux minimization, 30 PC & 7.84 & 449.01 & 210.21 \\
Flux minimization, 40 PC & 7.86 & 450.24 & 210.33 \\
MCMC sampling, 20 PC & $7.85^{+0.02}_{-0.02}$ & $449.74^{+1.30}_{-1.34}$ & $210.26^{+0.11}_{-0.11}$ \\
\citet{absil2013}, PCA & $8.01 \pm 0.16$ & $452 \pm 10$ & $211.2 \pm 1.3$ \\
\citet{cantalloube2015}, ANDROMEDA & $8.09 \pm 0.21$ & $455 \pm 8.7$ & $210.7 \pm 0.8$ \\
\hline
 & \multicolumn{3}{c}{VLT/NACO $M'$ filter} \\
Hessian minimization, 10 PC & 7.76 & 458.65 & 211.71 \\
Hessian minimization, 20 PC & 7.74 & 458.31 & 211.44 \\
Hessian minimization, 30 PC & 7.78 & 459.73 & 211.42 \\
Hessian minimization, 40 PC & 7.65 & 459.51 & 211.21 \\
Flux minimization, 10 PC & 7.63 & 456.95 & 211.61 \\
Flux minimization, 20 PC & 7.64 & 457.58 & 211.40 \\
Flux minimization, 30 PC & 7.65 & 459.31 & 211.40 \\
Flux minimization, 40 PC & 7.59 & 459.53 & 211.25 \\
MCMC sampling, 20 PC & $7.64^{+0.05}_{-0.05}$ & $458.42^{+3.53}_{-3.51}$ & $211.39^{+0.23}_{-0.24}$ \\
\citet{bonnefoy2013}, CADI & $7.5 \pm 0.3$ & & \\
\citet{bonnefoy2013}, KLIP & $7.8 \pm 0.3$ & & \\
\hline
\end{tabular}
\egroup
\end{table*}

Remaining bad pixels were removed with sigma clipping of outliers, similar to the $L'$ data. A frame selection was then applied to remove low-quality images by measuring the flux with a circular aperture (1~FWHM in radius) centered on the star. Frames for which the photometry deviated by more than $2\sigma$ from the median photometry were removed, corresponding to 4\% of the frames. Images were then centered with pixel precision by cropping around the brightest pixel. Next, a cross-correlation was used to align the images by shifting them to the mean offset of ten randomly drawn reference images. Finally, the stellar PSF was placed in the center of the frame with subpixel precision by fitting a 2D Gaussian to the mean of all images.

In preparation of the PSF subtraction, a pre-stacking by 100 images was applied and the central ($\le 1$~FWHM) and outer ($\ge 1\ffarcs0$) regions were masked. The stack of 522 images was decomposed into an orthogonal basis with the SVD, after which the model PSF was constructed by projecting each image onto the PCs in the range of 1--50. The derotated and median combined residuals obtained with 20~PCs are displayed in the bottom row of Fig.~\ref{fig:images} (second image from left). The computation of the S/N yielded a maximum of $\sim$23 with 30 PCs used for the PSF subtraction, approximately independent of the background-subtraction method.

\subsection{Photometry and astrometry of $\beta$~Pic~b}\label{sec:beta_pic_photometry}

The relative photometry and astrometry of $\beta$~Pic~b is determined with the two methods described in Sect.~\ref{sec:photastro}. For this purpose we used the pre-stacked images (594 images in $L'$ and 522 images in $M'$) and cropped both the science data and the unsaturated PSF templates (see first column in Fig.~\ref{fig:images}) to $2\arcsec \times 2\arcsec$.

The simplex minimization and MCMC analysis rely on the injection of negative artificial planets. With these methods, the PSF template of $\beta$~Pic was scaled by the relative brightness that was tested, the difference in DIT between the science data and the unsaturated exposures, and the ND filter that was used for the unsaturated images with the $M'$ filter. We adopted a filter transmission of $(2.33 \pm 0.10)$\% from \citet{bonnefoy2013}, which was measured on-sky for the combination of the \texttt{ND\_long} and $M'$ filter. Pixels at separations larger than $1\ffarcs0$ and smaller than 1.5~FWHM in $L'$ and 1.0~FWHM in $M'$ were masked after the artificial planet was injected and the stellar PSF was subsequently subtracted. This procedure was iterated for each step (i.e., different brightness and position) of the minimization and MCMC analysis.

\subsubsection{Simplex minimization of the residuals}\label{sec:beta_pic_simplex}

\begin{figure*}
\centering
\includegraphics[width=\linewidth]{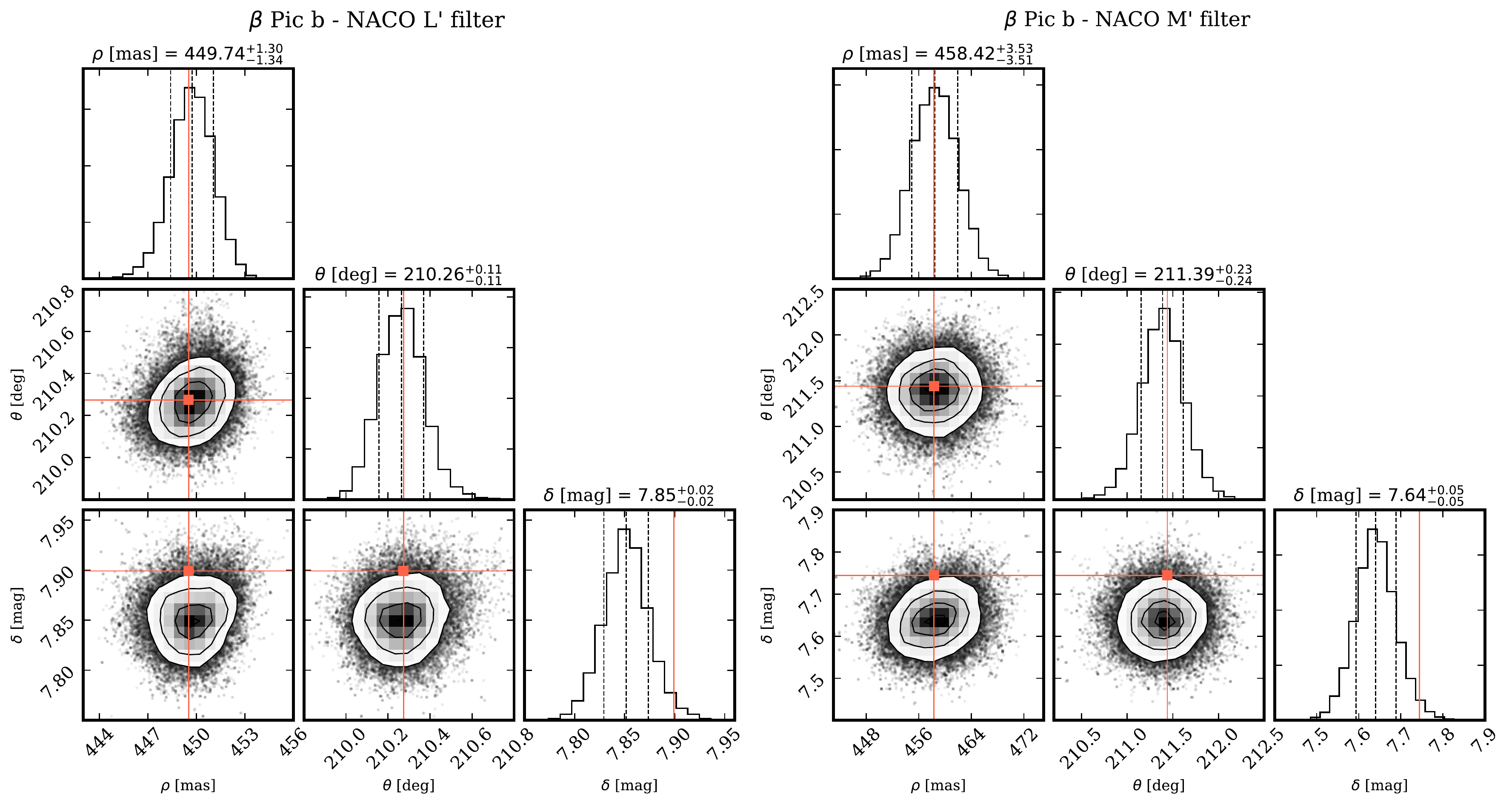}
\caption{Posterior probability distributions of the separation, $\rho$, position angle, $\theta$, and planet-to-star flux contrast, $\delta$, of $\beta$~Pic~b in $L'$ (\emph{left panel}) and $M'$ (\emph{right panel}). The marginalized 1D distributions of the parameters are shown in the \emph{diagonal} panels and the marginalized 2D distributions are shown for all parameter pairs in the \emph{off-axis} panels. The parameter values and uncertainties above the diagonal panels correspond to the median, and 16th and 84th percentiles of the samples (also indicated by the vertically dashed lines). Contours overlaid on the 2D distributions correspond to $1\sigma$, $2\sigma$, and $3\sigma$ confidence levels. The best-fit results from the Hessian minimization are shown with red symbols. See Sect.~\ref{sec:beta_pic_error} for a derivation of the photometric error budget.\label{fig:mcmc}}
\end{figure*}

The first analysis involved the minimization of the curvature or flux of the PSF subtraction residuals. From the initial guess, we let the simplex minimization iterate towards a best-fit solution with an absolute precision of 0.01~mag and 0.01~pixels. We smoothed the image residuals of the $L'$ and $M'$ data with a Gaussian filter of $\sigma = 30$~mas and $\sigma = 40$~mas ($\sim$1--1.5~pixels), respectively, in order to reduce pixel-to-pixel variations before the merit function of the Hessian was calculated (see Eq.~\ref{eq:hessian}). The standard deviation that was chosen for the kernel provided the highest accuracy, which has been quantified for both filters by measuring the position and brightness of artificial planets. The impact of the Gaussian filter on the photometric and astrometric precision is described and quantified in more detail in Sect.~\ref{sec:beta_pic_local}. The second merit function (see Eq.~\ref{eq:sum}) was calculated directly from the image residuals with no further smoothing applied. Both merit functions were evaluated within a circular aperture with a radius of 2~FWHM which was centered on the approximate and fixed position of $\beta$~Pic~b.

The results of the simplex minimization of the two merit functions are presented in Table~\ref{table:beta_pic_analysis} with 10, 20, 30, and 40 PCs used to model the PSF. The separations were calculated by assuming a pixel scale of 27.1~mas for both filters \citep[cf.][]{chauvin2012}. Literature values from \citet{absil2013,cantalloube2015,bonnefoy2013} are listed for reference. \citet{bonnefoy2013} minimized the standard deviation of the residuals and applied four different PSF-subtraction methods. In Table~\ref{table:beta_pic_analysis}, we only list their minimum and maximum contrast values. Several additional contrast values for the NACO $L'$ and $M'$ filters are available in the literature, which are derived from different data sets. These include $7.7 \pm 0.3$~mag \citep{lagrange2009}, $7.8 \pm 0.3$~mag \citep{lagrange2010}, $7.71 \pm 0.06$~mag \citep{currie2011}, and $7.79 \pm 0.08$~mag \citep{currie2013} in $L'$, and $7.50 \pm 0.13$~mag \citep{currie2013} in $M'$. We note that a precise, quantitative comparison of the photometric and astrometric values with the literature is challenging since small deviations are to be expected, for example due to the choice of the PSF template, centering method, and merit function. The photometric error budget is investigated in more detail in Sect.~\ref{sec:beta_pic_error}, but no astrometric calibration has been applied.

The dependence on the number of PCs that were tested appears small for the $L'$ data while there is some dispersion in the measured contrast values in the $M'$ data, in particular with the minimization of the Hessian. For the $M'$ data, the contrast was smallest with the largest number of PCs, both with the minimization of the Hessian and the flux. The values of the separation and position angle show differences of $\lesssim$1~mas and $\lesssim$0\ffdeg1 between the two minimization functions (both in the $L'$ and $M'$ filter) when the same number of PCs is compared. The contrast on the other hand is systematically smaller for both filters with the minimization of the flux, with the largest differences  ($\sim$0.1~mag) being seen in $M'$. This might be caused by a different impact of the residual speckle noise on the evaluation of the two merit functions. The best-fit residuals of the Hessian minimization (see third column in Fig.~\ref{fig:images}) show a relatively bright feature in the $L'$ and $M'$ data at the position of $\beta$~Pic~b. These noise features became more strongly suppressed by the flux minimization which resulted in a smaller contrast value. A quantification of the residual speckle noise and the impact on the photometric error budget are provided in Sect.~\ref{sec:beta_pic_error}.

\subsubsection{Markov chain Monte Carlo analysis}\label{sec:beta_pic_mcmc}

\begin{table*}
\caption{Photometry and error budget of $\beta$~Pic~b.}
\label{table:photometry}
\centering
\bgroup
\def\arraystretch{1.25}
\begin{tabular}{L{2.25cm} C{1.15cm} C{1.1cm} C{1.15cm} C{1.6cm} C{1.3cm} C{1.5cm} C{1.65cm} C{2.45cm}}
\hline\hline
Filter & MCMC contrast & MCMC error & Speckle error & Calibration error & ND filter error & Final contrast & Apparent magnitude & Apparent flux \\
 & (mag) & (mag) & (mag) & (mag) & (mag) & (mag) & (mag) & (W m$^{-2}$ $\mu$m$^{-1}$) \\
\hline
VLT/NACO $L'$ & 7.85 & 0.02 & 0.03 & 0.05 & -    & $7.85 \pm 0.06$ & $11.30 \pm 0.06$ & $1.53 \pm 0.08 \times 10^{-15}$\\
VLT/NACO $M'$ & 7.64 & 0.05 & 0.04 & 0.09 & 0.05 & $7.64 \pm 0.12$ & $11.10 \pm 0.12$ & $7.63 \pm 0.84 \times 10^{-16}$\\
\hline
\end{tabular}
\egroup
\end{table*}

The correlations between the separation, position angle, and contrast of $\beta$~Pic~b, as well as the statistical uncertainties related to the Poisson noise were estimated with MCMC sampling (see Sect.~\ref{sec:photastro}). The same stack of images, mask, and aperture were used such that a robust comparison can be made with the results from the simplex minimization. The probability landscape was explored by 200 walkers, each one of them creating a chain of 500 steps. The mean acceptance fraction of the samples was 0.64 in $L'$ and 0.66 in $M'$. The integrated autocorrelation time of the samples was estimated to be $\tau_\mathrm{int} = {38.1, 1.3, 23.3}$ in $L'$ and $\tau_\mathrm{int} = {32.7, 38.0, 30.1}$ in $M'$ with the listed values for the time series of the separation, position angle, and contrast. The chains can be considered converged if we assume that their length should exceed $10\tau_\mathrm{int}$ for all parameters. The first 50 samples of the chains were identified upon visual inspection as the burn-in so they were excluded from the analysis. The marginalized posterior distributions and the derived uncertainties on the parameters are presented in Fig.~\ref{fig:mcmc}, and listed in Table~\ref{table:beta_pic_analysis} in comparison with the results from the simplex minimization.

The best-fit values from the minimization of the flux residuals with 20~PCs are very similar to the median values of the posterior distributions estimated with the MCMC analysis. The separations and position angles that are retrieved with the Hessian minimization are within the $1\sigma$ uncertainties that are derived from the posterior distributions. The contrast on the other hand obtained with the Hessian minimization and the same number of PCs is systematically slightly higher compared to the results from the MCMC analysis. The values deviate by $2.5\sigma$ ($L'$) and $2\sigma$ ($M'$) from the MCMC results, with a difference of 0.05~mag in $L'$ and 0.1~mag in $M'$ (see Fig.~\ref{fig:mcmc} and Table~\ref{table:beta_pic_analysis}). As mentioned in Sect.~\ref{sec:beta_pic_simplex}, this discrepancy is likely related to the residual speckle noise which biases the two merit functions differently.

\subsubsection{Photometric error budget}\label{sec:beta_pic_error}

The uncertainties presented in Fig.~\ref{fig:mcmc} are derived from the posterior distributions alone, therefore they only reflect the statistical error and not the total error budget. Here we provide more realistic error bars on the photometry of $\beta$~Pic~b in $L'$ and $M'$ by considering four additional terms in the error budget. Firstly, we estimated the residual speckle noise which biases the calculation of the likelihood function as described in more detail below. Secondly, there is a calibration error related to the brightness of the star which varied during the observations due to changes in Strehl ratio and sky conditions while the same PSF template is injected in all images. This error was calculated as the standard deviation of the stellar flux in the unsaturated images, measured with a circular aperture of 1.5~FWHM in radius. Thirdly, we included the error on the transmission of the ND filter for which we adopted an absolute error of 0.10\% from \citet{bonnefoy2013}. Finally, we considered the uncertainty on the apparent magnitude of $\beta$~Pic in the $L'$ and $M'$ filters.

The uncertainty caused by the residual speckle noise was estimated with the procedure described in \citet{wertz2017}. We briefly summarize that an artificial planet was injected at the separation and with the brightness of $\beta$~Pic~b (which was first removed with the best-fit solution), after which its position and brightness were retrieved by minimizing the flux residuals (see Eq.~\ref{eq:sum}). This procedure was then repeated at 360 equally spaced position angles. The distributions of the offsets between the injected and retrieved values of the separation, position angle, and flux contrast were fitted with a Gaussian function and evaluated with a maximum-likelihood estimation. The best-fit mean and standard deviation (in parentheses) of the three parameter offsets are $3.6\times10^{-2}$ $(9.4\times10^{-1})$~mas, $1\ffdeg5\times10^{-3}$ $(1\ffdeg2\times10^{-1})$, and $9.1\times10^{-4}$ $(3.3\times10^{-2})$~mag in $L'$, and $7.2\times10^{-2}$ $(1.6\times10^{0})$~mas, $3\ffdeg8\times10^{-5}$ $(1\ffdeg8\times10^{-1})$, and $1.0\times10^{-3}$ $(4.0\times10^{-2})$~mag in $M'$. There appears no systematic uncertainty as the mean values are all consistent with zero so we only consider the width of the distributions.

\begin{figure}
\centering
\includegraphics[width=\linewidth]{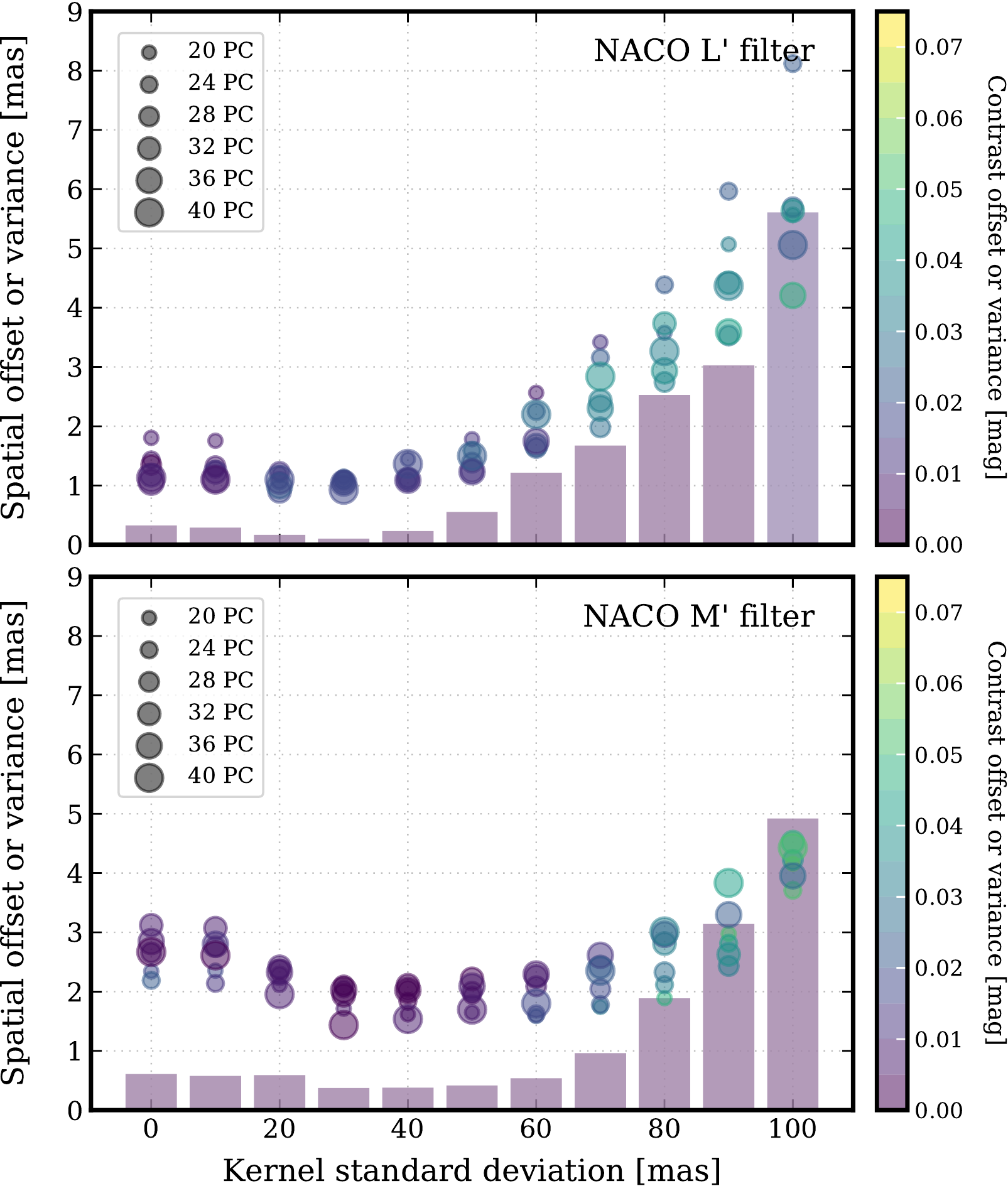}
\caption{Photometric and astrometric precision obtained with the minimization of the Hessian for the $L'$ (\textit{top panel}) and $M'$ (\textit{bottom panel}) data. The circular points show the azimuthally averaged position and contrast offset between the injected and retrieved values. The symbol sizes increase in steps of 4~PCs in the range of 20--40~PCs. The precision is tested for different widths (increasing in steps of 10~mas) of the Gaussian filter that is used to lower pixel-to-pixel variations before the Hessian is calculated. The bars show the variance of the six azimuthal positions, averaged over the six values of the PCs that were tested. The color coding is the same for the bars and the data points. \label{fig:simplex}}
\end{figure}

The error components of the photometry are independent and combined as the square root of the sum of the squares. Table~\ref{table:photometry} lists the individual errors for both filters, as well as the final contrast with uncertainty. The apparent magnitude of $\beta$~Pic is $3.454 \pm 0.003$~mag in $L'$ and $3.458 \pm 0.009$~mag in $M'$ \citep{bouchet1991}. The uncertainty on the stellar magnitude is included in the error budget of $\beta$~Pic~b but it is negligible for the uncertainty on the planet's apparent magnitude. Therefore, this term is not listed in Table~\ref{table:photometry}. The flux loss related to the off-axis throughput of the AGPM coronagraph \citep{mawet2013} is negligible at the location of $\beta$~Pic~b. The zero-point flux for the two filters is computed by folding a flux-calibrated spectrum of Vega \citep{bohlin2007} with the filter transmission\footnote{\url{http://www.eso.org/sci/facilities/paranal/instruments/naco/inst/filters.html}}. The apparent fluxes of $\beta$~Pic~b (see last column in Table~\ref{table:photometry}) are derived by setting the magnitude of Vega to zero for each filter.

\begin{figure*}
\centering
\includegraphics[width=0.9\linewidth]{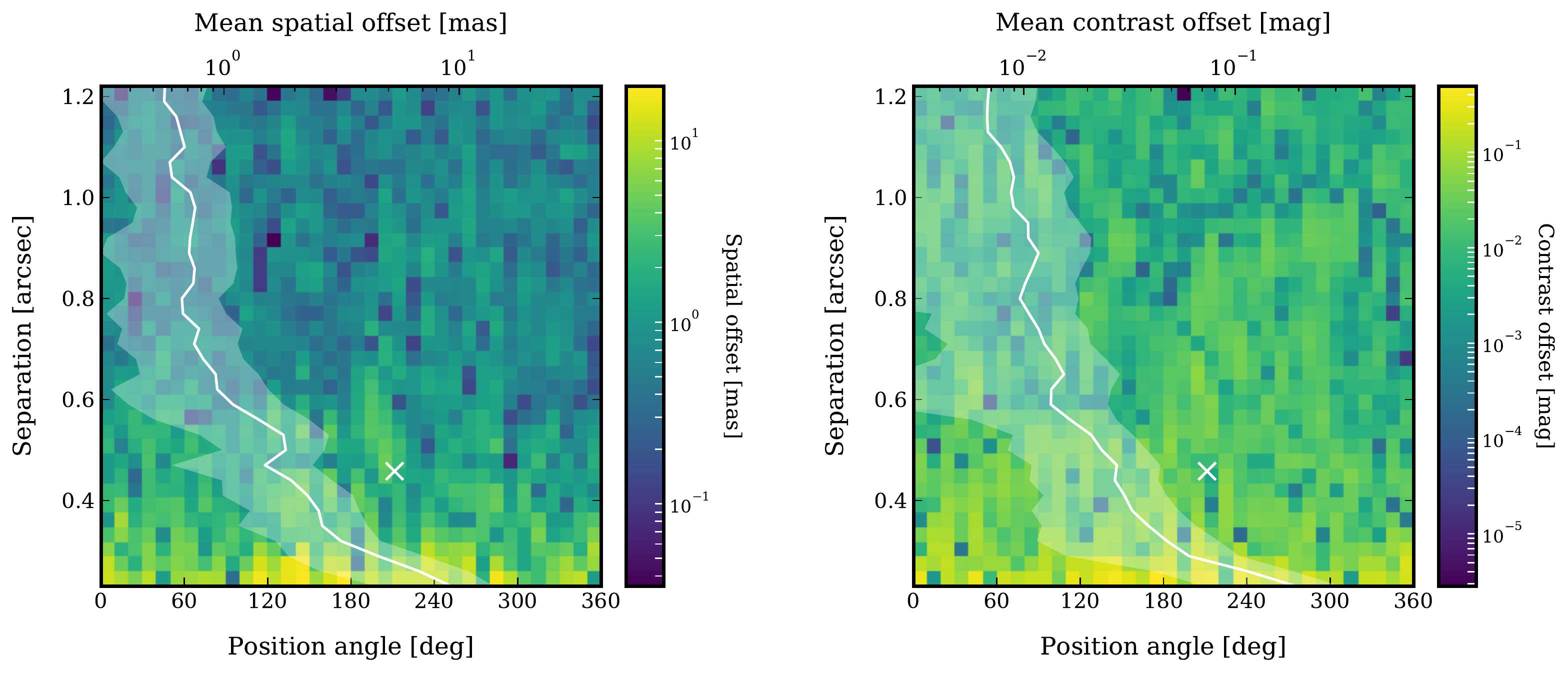}
\caption{Locality of the photometric and astrometric precision tested with the Hessian minimization. Artificial planets are injected in the $M'$ filter data at a range of separations and position angles. The absolute offsets between the injected and retrieved value of the position (\textit{left panel}) and the contrast (\textit{right panel}) are shown on a logarithmic color scale for each tested position. The white solid line is the mean offset (top axis) as a function of separation (left axis). The standard deviation of the offset across the position angles is indicated by the white shaded area. The white cross denotes the position of $\beta$~Pic~b, which was removed beforehand.\label{fig:local}}
\end{figure*}

As we are mostly interested in measuring MIR photometry with related error bars, no astrometric calibration is performed on the data and the error budget of the planet's position is not investigated. The $M'$ filter in particular is not well suited to high-precision astrometry. The position angles listed in Table~\ref{table:photometry} have therefore not been corrected for true north. Details on the instrumental uncertainties of NACO can for instance be found in the work by \citet{rameau2013} and \citet{chauvin2012,chauvin2015}. For example, \citet{chauvin2012} measured a pixel scale of $27.11 \pm 0.04$~mas and a true north offset of $-0\ffdeg36 \pm 0\ffdeg11$ with NACO in the $L'$ filter.

\subsubsection{Effects of smoothing and spatial variations}\label{sec:beta_pic_local}

As described in Sect.~\ref{sec:beta_pic_simplex}, singular residuals are smoothed with a Gaussian filter to lower the impact of pixel-to-pixel variations on the Hessian. Here we investigate the dependence of the smoothing on the photometric and astrometric precision. We started again with a data cube from which $\beta$~Pic~b had been removed and injected an artificial planet with a similar brightness and separation as $\beta$~Pic~b. We then applied the Hessian minimization (see Eq.~\ref{eq:hessian}) and compared the retrieved position and brightness with the injected values. The calculation was repeated for different numbers of PCs ranging from 20 to 40, six equally spaced position angles, and different widths of the Gaussian filter.

The results are presented in Fig.~\ref{fig:simplex} for both data sets. For the $L'$ data, the azimuthally averaged offset between the measured and injected position values shows a minimum around 30~mas while the photometric precision is largest if no smoothing is applied. Also, the variance of the six azimuthal positions, averaged over the different PCs that were tested, shows a minimum around a similar width of the Gaussian filter. Therefore, smoothing provides a small improvement to the astrometric accuracy ($\lesssim$1.1~mas) with the Hessian minimization if instrumental calibration errors are neglected. However, we note that the impact of smoothing on the precision will likely depend on the amount of residual noise at a given separation. The same analysis for the $M'$ data shows a minimum in both the spatial and contrast offset when a Gaussian standard deviation of approximately 40~mas is used. For the separation and brightness of $\beta$~Pic~b, the derived astrometric and photometric precision in $M'$ is $\lesssim$2.5~mas and $\lesssim$0.01~mag, respectively.

The variance that is shown in Fig.~\ref{fig:simplex} indicates that the measurement precision is not circular symmetric but affected by local variations in noise residuals. In Sect.~\ref{sec:beta_pic_error}, we estimated the uncertainty on the photometry of $\beta$~Pic~b caused by residual speckle noise. Here we extend that analysis with a quantification of localized variations both in radial and azimuthal directions with the Hessian minimization. The precision is tested across a grid of 34 equally spaced separations in the range of $0\ffarcs23$--$1\ffarcs19$ and 36 equally spaced position angles while fixing the number of PCs to 20 and the Gaussian filter width to 40~mas. The result is visualized in Fig.~\ref{fig:local}, which shows the absolute spatial and contrast offset between the injected and retrieved values. Overall, both the photometric and astrometric precision increases towards larger separations as the amount of noise residuals decreases. In addition, asymmetric variations appear on various spatial scales, indicating that photometric and astrometric measurements are a localized problem.

\subsection{Detection limits}\label{sec:beta_pic_limits}

Detection limits were calculated by injecting a copy of the unsaturated stellar PSF at a range of separations and position angles, as described in Sect.~\ref{sec:detection_limits}. The brightness of the artificial planet was iteratively adjusted until the FPF converged to a fixed $5\sigma$ level. This implies that the FPF increases towards smaller separations, following the small sample statistics \citep{mawet2014}. A fractional tolerance on the FPF of 0.1 was chosen to end the iteration process. The FPF was calculated by placing a circular aperture at the position of the artificial planet and filling the remaining azimuthal space with nonoverlapping reference apertures to estimate the noise level. The apertures next to the planet were neglected because they encircled the self-subtraction lobes of the planet, otherwise biasing the noise measurement (see fourth column in Fig.~\ref{fig:images} for an example). We chose a conservative aperture diameter equal to the FWHM of the stellar PSF which is sufficiently large for the noise samples to be independent. Detection limits were calculated for PCs in the range of 10--50 with steps of 5~PCs. No correction was applied for the off-axis throughput of the coronagraph since the flux loss beyond 300~mas is less than 0.1\% \citep{mawet2013}.

\begin{figure}
\centering
\includegraphics[width=\linewidth]{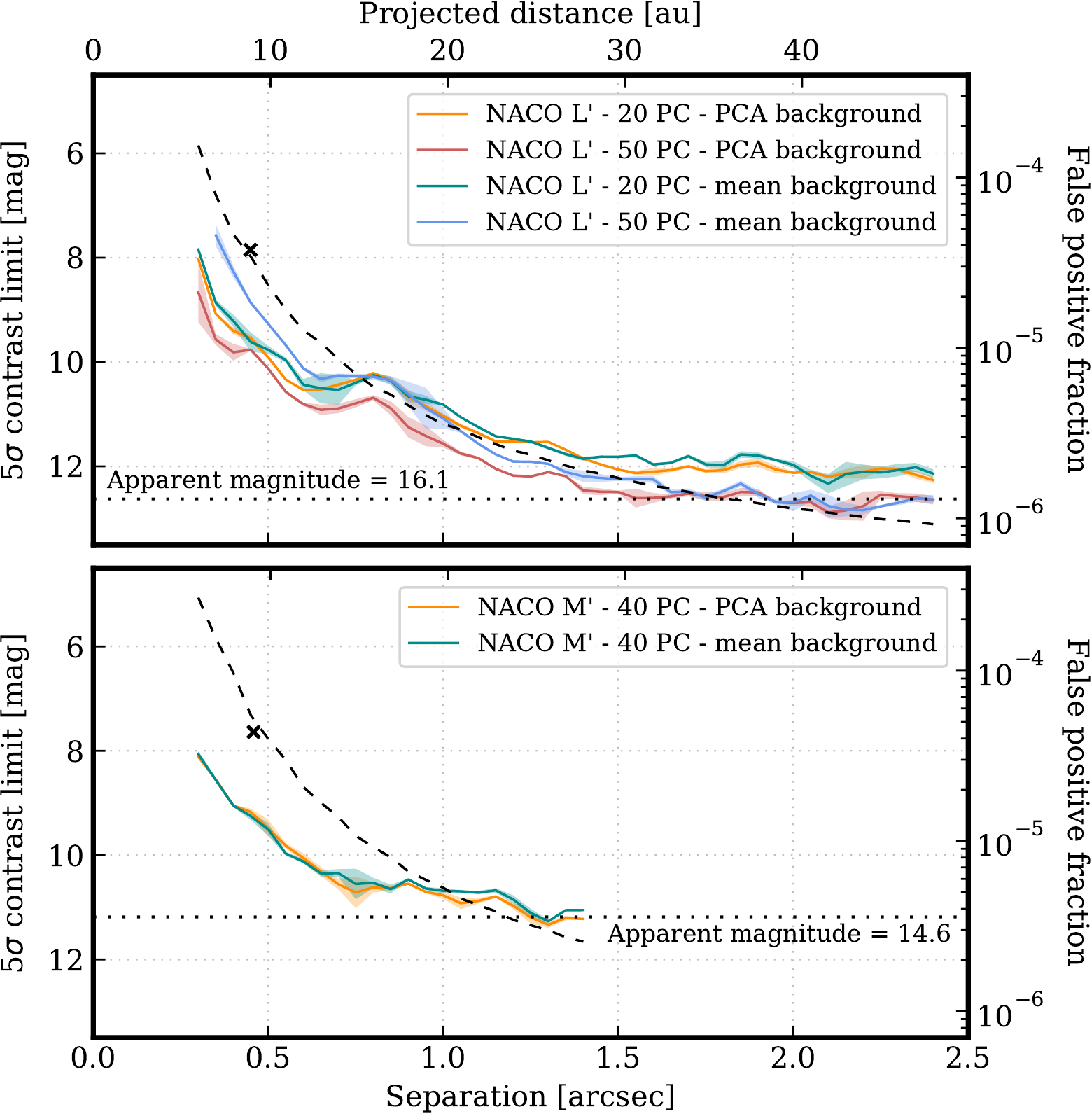}
\caption{Detection limits in the NACO $L'$ (\textit{top panel}) and $M'$ (\textit{bottom panel}) filters. Solid lines are the azimuthally averaged limits and the shaded areas are the variance of the six azimuthal positions. The limits are calculated with a mean and PCA-based background subtraction for both filters. A range of PCs was tested but only the limits with the highest sensitivity are presented. The black dashed lines show the false positive fraction (right axis) associated with the detection limits and the horizontally dotted lines indicate the apparent magnitude in the background-limited regime. The black crosses show the separation and contrast of $\beta$~Pic~b as determined with the MCMC analysis (the uncertainties are smaller than the symbol size).\label{fig:limits}}
\end{figure}

Figure~\ref{fig:limits} shows the $5\sigma$ detection limits with radial step sizes of 50~mas for both NACO filters, averaged over six equally spaced azimuthal positions. The FPF associated with the $\sigma$ level and number of reference apertures is also shown for reference. The Gaia DR2 distance of $19.75 \pm 0.13$~pc \citep{gaia2016,gaia2018} is adopted to determine the projected distance from the star. The background-limited regime starts approximately at a separation of 1\ffarcs5 in $L'$ and 1\ffarcs2 in $M'$, as estimated by eye from the flattening of the contrast curves. The limiting apparent magnitude in this regime is approximately 16.1~mag in $L'$ and 14.7~mag in $M'$ which is calculated by averaging the contrast limits over the background-limited separations for the most sensitive number of PCs. We note that the total amount of data that was used (after frame selection) corresponds to an integration time of 99~min in the $L'$ filter and 56~min in the $M'$ filter.

As described in Sects.~\ref{sec:naco_lp} and \ref{sec:naco_mp}, both data sets were processed with a mean- and PCA-based background subtraction separately while all other pipeline modules were executed in the same way. To investigate the impact of both approaches on the sensitivity, we calculated the detection limits for both cases with the same PSF template. Furthermore, the dependence of the detection limits on the number of PCs was determined by visually inspecting the contrast curves and selecting those with the best limits.

In the $L'$ filter, there is a difference between the detection limits for the two background subtraction methods. With the mean background subtraction, the highest sensitivity was reached with 20~PCs in the speckle-limited regime and 50~PCs in the background-limited regime. A smaller number of PCs is sufficient at smaller separations as the limiting contrast decreases and the effect of self-subtraction increases. Interestingly, with the PCA-based background subtraction the best limits were obtained with 50~PCs both at small and large separations. In the $M'$ filter on the other hand, the dependence of the detection limits on the number of PCs is small (up to the maximum separation that was probed by the observations) with the best limits overall being obtained with 40~PCs, independent of the background subtraction method.

\section{Summary}\label{sec:summary}

We have presented the new pipeline architecture of PynPoint: an open-source Python package for processing and analysis of high-contrast imaging data. The package provides a generic, end-to-end data-reduction pipeline, including analysis tools. The architecture of the pipeline has a modular design with the core functionalities and the pipeline modules separately implemented, which ensures scalability to new data formats and pipeline modules while the robustness of the pipeline remains secured. Dedicated pipeline modules have been implemented to import and export data, and to process and analyze data sets, including modules for background subtraction, frame registration, bad-pixel cleaning (including the corrections of bad-pixel clusters using spectral deconvolution), PSF subtraction with full-frame PCA, estimation of detection limits, and photometric and astrometric measurements of companions.

Reprocessing and analysis of archival VLT/NACO data of $\beta$~Pictoris demonstrates the applicability of the pipeline to MIR data with dedicated background subtraction modules for data obtained with dithering or nodding without having to pre-stack images. We determined the photometry and astrometry of $\beta$~Pic~b by injecting artificial planets and using a minimization algorithm and MCMC analysis. Minimization of the curvature of the PSF-subtraction residuals yielded contrast values that were up to $\sim$0.1~mag larger compared to the minimization of the flux of the residuals. This inconsistency is probably caused by a different impact of the residual speckle noise on the evaluation of the two merit functions. The final contrast values and uncertainties of $\beta$~Pic~b in the NACO $L'$ and $M'$ filters are $7.85 \pm 0.06$~mag and $7.64 \pm 0.12$~mag, respectively. The photometric error includes the statistical uncertainty, the residual speckle noise, the calibration error, and the transmission uncertainty of the ND filter.

PynPoint is under continuous development and we welcome contributions from the high-contrast imaging community to help extend and improve the pipeline. The architecture with its abstract interface and pipeline functionalities allows for easy implementation of new pipeline modules or improvements of existing ones (more details are provided in the online documentation). PynPoint is maintained on Github\footnote{\url{https://github.com/PynPoint/PynPoint}} and is also available in the PyPI repository\footnote{\url{https://pypi.org/project/pynpoint/}}.

\begin{acknowledgements}

We thank Silvan Hunziker, Janis Hagelberg, and Jonas K\"{u}hn for valuable discussions, Dimitri Mawet for sharing the transmission profile of the AGPM coronagraph, and the referee for providing constructive comments that helped to improve the quality of this paper. TS acknowledges the support from the ETH Zurich Postdoctoral Fellowship Program. SPQ thanks the Swiss National Science Foundation (SNSF) for financial support under grant number 200021\_169131. Part of this work was carried out within the framework of the National Centre for Competence in Research PlanetS supported by the Swiss National Science Foundation. SPQ acknowledges the financial support of the SNSF.

\end{acknowledgements}

\bibliographystyle{aa}
\bibliography{references}

\begin{thebibliography}{68}
\expandafter\ifx\csname natexlab\endcsname\relax\def\natexlab#1{#1}\fi

\bibitem[{Aach \& Metzler(2001)}]{aach2001}
Aach, T. \& Metzler, V.~H. 2001, in Medical Imaging 2001: Image Processing,
  Vol. 4322, International Society for Optics and Photonics, 824--836

\bibitem[{{Absil} {et~al.}(2013){Absil}, {Milli}, {Mawet}, {Lagrange},
  {Girard}, {Chauvin}, {Boccaletti}, {Delacroix}, \& {Surdej}}]{absil2013}
{Absil}, O., {Milli}, J., {Mawet}, D., {et~al.} 2013, \aap, 559, L12

\bibitem[{{Amara} \& {Quanz}(2012)}]{amara2012}
{Amara}, A. \& {Quanz}, S.~P. 2012, \mnras, 427, 948

\bibitem[{{Amara} {et~al.}(2015){Amara}, {Quanz}, \& {Akeret}}]{amara2015}
{Amara}, A., {Quanz}, S.~P., \& {Akeret}, J. 2015, Astronomy and Computing, 10,
  107

\bibitem[{{Barman} {et~al.}(2015){Barman}, {Konopacky}, {Macintosh}, \&
  {Marois}}]{barman2015}
{Barman}, T.~S., {Konopacky}, Q.~M., {Macintosh}, B., \& {Marois}, C. 2015,
  \apj, 804, 61

\bibitem[{{Beuzit} {et~al.}(2008){Beuzit}, {Feldt}, {Dohlen}, {Mouillet},
  {Puget}, {Wildi}, {Abe}, {Antichi}, {Baruffolo}, {Baudoz}, {Boccaletti},
  {Carbillet}, {Charton}, {Claudi}, {Downing}, {Fabron}, {Feautrier},
  {Fedrigo}, {Fusco}, {Gach}, {Gratton}, {Henning}, {Hubin}, {Joos}, {Kasper},
  {Langlois}", {Lenzen}, {Moutou}, {Pavlov}, {Petit}, {Pragt}, {Rabou},
  {Rigal}, {Roelfsema}, {Rousset}, {Saisse}, {Schmid}, {Stadler}, {Thalmann},
  {Turatto}, {Udry}, {Vakili}, \& {Waters}}]{beuzit2008}
{Beuzit}, J.-L., {Feldt}, M., {Dohlen}, K., {et~al.} 2008, in Society of
  Photo-Optical Instrumentation Engineers (SPIE) Conference Series, Vol. 7014,
  Society of Photo-Optical Instrumentation Engineers (SPIE) Conference Series,
  18

\bibitem[{{Biller} {et~al.}(2013){Biller}, {Liu}, {Wahhaj}, {Nielsen},
  {Hayward}, {Males}, {Skemer}, {Close}, {Chun}, {Ftaclas}, {Clarke}, {Thatte},
  {Shkolnik}, {Reid}, {Hartung}, {Boss}, {Lin}, {Alencar}, {de Gouveia Dal
  Pino}, {Gregorio-Hetem}, \& {Toomey}}]{biller2013}
{Biller}, B.~A., {Liu}, M.~C., {Wahhaj}, Z., {et~al.} 2013, \apj, 777, 160

\bibitem[{{Bohlin}(2007)}]{bohlin2007}
{Bohlin}, R.~C. 2007, in Astronomical Society of the Pacific Conference Series,
  Vol. 364, The Future of Photometric, Spectrophotometric and Polarimetric
  Standardization, ed. C.~{Sterken}, 315

\bibitem[{{Bonnefoy} {et~al.}(2013){Bonnefoy}, {Boccaletti}, {Lagrange},
  {Allard}, {Mordasini}, {Beust}, {Chauvin}, {Girard}, {Homeier}, {Apai},
  {Lacour}, \& {Rouan}}]{bonnefoy2013}
{Bonnefoy}, M., {Boccaletti}, A., {Lagrange}, A.-M., {et~al.} 2013, \aap, 555,
  A107

\bibitem[{{Bonse} {et~al.}(2018){Bonse}, {Quanz}, \& {Amara}}]{bonse2018}
{Bonse}, M.~J., {Quanz}, S.~P., \& {Amara}, A. 2018, ArXiv e-prints
  [\eprint[arXiv]{1804.05063}]

\bibitem[{{Bouchet} {et~al.}(1991){Bouchet}, {Manfroid}, \&
  {Schmider}}]{bouchet1991}
{Bouchet}, P., {Manfroid}, J., \& {Schmider}, F.~X. 1991, \aaps, 91, 409

\bibitem[{{Bowler}(2016)}]{bowler2016}
{Bowler}, B.~P. 2016, \pasp, 128, 102001

\bibitem[{{Brandt} {et~al.}(2013){Brandt}, {McElwain}, {Turner}, {Abe},
  {Brandner}, {Carson}, {Egner}, {Feldt}, {Golota}, {Goto}, {Grady}, {Guyon},
  {Hashimoto}, {Hayano}, {Hayashi}, {Hayashi}, {Henning}, {Hodapp}, {Ishii},
  {Iye}, {Janson}, {Kandori}, {Knapp}, {Kudo}, {Kusakabe}, {Kuzuhara}, {Kwon},
  {Matsuo}, {Miyama}, {Morino}, {Moro-Mart{\'{\i}}n}, {Nishimura}, {Pyo},
  {Serabyn}, {Suto}, {Suzuki}, {Takami}, {Takato}, {Terada}, {Thalmann},
  {Tomono}, {Watanabe}, {Wisniewski}, {Yamada}, {Takami}, {Usuda}, \&
  {Tamura}}]{brandt2013}
{Brandt}, T.~D., {McElwain}, M.~W., {Turner}, E.~L., {et~al.} 2013, \apj, 764,
  183

\bibitem[{{Brandt} {et~al.}(2014){Brandt}, {McElwain}, {Turner}, {Mede},
  {Spiegel}, {Kuzuhara}, {Schlieder}, {Wisniewski}, {Abe}, {Biller},
  {Brandner}, {Carson}, {Currie}, {Egner}, {Feldt}, {Golota}, {Goto}, {Grady},
  {Guyon}, {Hashimoto}, {Hayano}, {Hayashi}, {Hayashi}, {Henning}, {Hodapp},
  {Inutsuka}, {Ishii}, {Iye}, {Janson}, {Kandori}, {Knapp}, {Kudo}, {Kusakabe},
  {Kwon}, {Matsuo}, {Miyama}, {Morino}, {Moro-Mart{\'{\i}}n}, {Nishimura},
  {Pyo}, {Serabyn}, {Suto}, {Suzuki}, {Takami}, {Takato}, {Terada}, {Thalmann},
  {Tomono}, {Watanabe}, {Yamada}, {Takami}, {Usuda}, \& {Tamura}}]{brandt2014}
{Brandt}, T.~D., {McElwain}, M.~W., {Turner}, E.~L., {et~al.} 2014, \apj, 794,
  159

\bibitem[{{Buitinck} {et~al.}(2013){Buitinck}, {Louppe}, {Blondel},
  {Pedregosa}, {Mueller}, {Grisel}, {Niculae}, {Prettenhofer}, {Gramfort},
  {Grobler}, {Layton}, {Vanderplas}, {Joly}, {Holt}, \&
  {Varoquaux}}]{scikit-learn2013}
{Buitinck}, L., {Louppe}, G., {Blondel}, M., {et~al.} 2013, ArXiv e-prints
  [\eprint[arXiv]{1309.0238}]

\bibitem[{{Burrows} {et~al.}(1997){Burrows}, {Marley}, {Hubbard}, {Lunine},
  {Guillot}, {Saumon}, {Freedman}, {Sudarsky}, \& {Sharp}}]{burrows1997}
{Burrows}, A., {Marley}, M., {Hubbard}, W.~B., {et~al.} 1997, \apj, 491, 856

\bibitem[{{Cantalloube} {et~al.}(2015){Cantalloube}, {Mouillet}, {Mugnier},
  {Milli}, {Absil}, {Gomez Gonzalez}, {Chauvin}, {Beuzit}, \&
  {Cornia}}]{cantalloube2015}
{Cantalloube}, F., {Mouillet}, D., {Mugnier}, L.~M., {et~al.} 2015, \aap, 582,
  A89

\bibitem[{{Chauvin} {et~al.}(2017){Chauvin}, {Desidera}, {Lagrange}, {Vigan},
  {Gratton}, {Langlois}, {Bonnefoy}, {Beuzit}, {Feldt}, {Mouillet}, {Meyer},
  {Cheetham}, {Biller}, {Boccaletti}, {D'Orazi}, {Galicher}, {Hagelberg},
  {Maire}, {Mesa}, {Olofsson}, {Samland}, {Schmidt}, {Sissa}, {Bonavita},
  {Charnay}, {Cudel}, {Daemgen}, {Delorme}, {Janin-Potiron}, {Janson},
  {Keppler}, {Le Coroller}, {Ligi}, {Marleau}, {Messina}, {Molli{\`e}re},
  {Mordasini}, {M{\"u}ller}, {Peretti}, {Perrot}, {Rodet}, {Rouan}, {Zurlo},
  {Dominik}, {Henning}, {Menard}, {Schmid}, {Turatto}, {Udry}, {Vakili}, {Abe},
  {Antichi}, {Baruffolo}, {Baudoz}, {Baudrand}, {Blanchard}, {Bazzon}, {Buey},
  {Carbillet}, {Carle}, {Charton}, {Cascone}, {Claudi}, {Costille}, {Deboulbe},
  {De Caprio}, {Dohlen}, {Fantinel}, {Feautrier}, {Fusco}, {Gigan}, {Giro},
  {Gisler}, {Gluck}, {Hubin}, {Hugot}, {Jaquet}, {Kasper}, {Madec}, {Magnard},
  {Martinez}, {Maurel}, {Le Mignant}, {M{\"o}ller-Nilsson}, {Llored}, {Moulin},
  {Orign{\'e}}, {Pavlov}, {Perret}, {Petit}, {Pragt}, {Puget}, {Rabou},
  {Ramos}, {Rigal}, {Rochat}, {Roelfsema}, {Rousset}, {Roux}, {Salasnich},
  {Sauvage}, {Sevin}, {Soenke}, {Stadler}, {Suarez}, {Weber}, {Wildi},
  {Antoniucci}, {Augereau}, {Baudino}, {Brandner}, {Engler}, {Girard}, {Gry},
  {Kral}, {Kopytova}, {Lagadec}, {Milli}, {Moutou}, {Schlieder},
  {Szul{\'a}gyi}, {Thalmann}, \& {Wahhaj}}]{chauvin2017}
{Chauvin}, G., {Desidera}, S., {Lagrange}, A.-M., {et~al.} 2017, \aap, 605, L9

\bibitem[{{Chauvin} {et~al.}(2012){Chauvin}, {Lagrange}, {Beust}, {Bonnefoy},
  {Boccaletti}, {Apai}, {Allard}, {Ehrenreich}, {Girard}, {Mouillet}, \&
  {Rouan}}]{chauvin2012}
{Chauvin}, G., {Lagrange}, A.-M., {Beust}, H., {et~al.} 2012, \aap, 542, A41

\bibitem[{{Chauvin} {et~al.}(2015){Chauvin}, {Vigan}, {Bonnefoy}, {Desidera},
  {Bonavita}, {Mesa}, {Boccaletti}, {Buenzli}, {Carson}, {Delorme},
  {Hagelberg}, {Montagnier}, {Mordasini}, {Quanz}, {Segransan}, {Thalmann},
  {Beuzit}, {Biller}, {Covino}, {Feldt}, {Girard}, {Gratton}, {Henning},
  {Kasper}, {Lagrange}, {Messina}, {Meyer}, {Mouillet}, {Moutou}, {Reggiani},
  {Schlieder}, \& {Zurlo}}]{chauvin2015}
{Chauvin}, G., {Vigan}, A., {Bonnefoy}, M., {et~al.} 2015, \aap, 573, A127

\bibitem[{{Crepp} {et~al.}(2011){Crepp}, {Pueyo}, {Brenner}, {Oppenheimer},
  {Zimmerman}, {Hinkley}, {Parry}, {King}, {Vasisht}, {Beichman},
  {Hillenbrand}, {Dekany}, {Shao}, {Burruss}, {Roberts}, {Bouchez}, {Roberts},
  \& {Soummer}}]{crepp2011}
{Crepp}, J.~R., {Pueyo}, L., {Brenner}, D., {et~al.} 2011, \apj, 729, 132

\bibitem[{{Currie} {et~al.}(2013){Currie}, {Burrows}, {Madhusudhan},
  {Fukagawa}, {Girard}, {Dawson}, {Murray-Clay}, {Kenyon}, {Kuchner},
  {Matsumura}, {Jayawardhana}, {Chambers}, \& {Bromley}}]{currie2013}
{Currie}, T., {Burrows}, A., {Madhusudhan}, N., {et~al.} 2013, \apj, 776, 15

\bibitem[{{Currie} {et~al.}(2011){Currie}, {Thalmann}, {Matsumura},
  {Madhusudhan}, {Burrows}, \& {Kuchner}}]{currie2011}
{Currie}, T., {Thalmann}, C., {Matsumura}, S., {et~al.} 2011, \apjl, 736, L33

\bibitem[{{Foreman-Mackey} {et~al.}(2013){Foreman-Mackey}, {Hogg}, {Lang}, \&
  {Goodman}}]{foreman2013}
{Foreman-Mackey}, D., {Hogg}, D.~W., {Lang}, D., \& {Goodman}, J. 2013, \pasp,
  125, 306

\bibitem[{Franke(1987)}]{franke1987}
Franke, U. 1987, in ICASSP '87. IEEE International Conference on Acoustics,
  Speech, and Signal Processing, Vol.~12, 1300--1303

\bibitem[{{Gaia Collaboration} {et~al.}(2018){Gaia Collaboration}, {Brown},
  {Vallenari}, {Prusti}, {de Bruijne}, {Babusiaux}, {Bailer-Jones}, {Biermann},
  {Evans}, {Eyer}, \& et~al.}]{gaia2018}
{Gaia Collaboration}, {Brown}, A.~G.~A., {Vallenari}, A., {et~al.} 2018, \aap,
  616, A1

\bibitem[{{Gaia Collaboration} {et~al.}(2016){Gaia Collaboration}, {Prusti},
  {de Bruijne}, {Brown}, {Vallenari}, {Babusiaux}, {Bailer-Jones}, {Bastian},
  {Biermann}, {Evans}, \& et~al.}]{gaia2016}
{Gaia Collaboration}, {Prusti}, T., {de Bruijne}, J.~H.~J., {et~al.} 2016,
  \aap, 595, A1

\bibitem[{{Galicher} {et~al.}(2018){Galicher}, {Boccaletti}, {Mesa}, {Delorme},
  {Gratton}, {Langlois}, {Lagrange}, {Maire}, {Le Coroller}, {Chauvin},
  {Biller}, {Cantalloube}, {Janson}, {Lagadec}, {Meunier}, {Vigan},
  {Hagelberg}, {Bonnefoy}, {Zurlo}, {Rocha}, {Maurel}, {Jaquet}, {Buey}, \&
  {Weber}}]{galicher2018}
{Galicher}, R., {Boccaletti}, A., {Mesa}, D., {et~al.} 2018, \aap, 615, A92

\bibitem[{{Galicher} {et~al.}(2016){Galicher}, {Marois}, {Macintosh},
  {Zuckerman}, {Barman}, {Konopacky}, {Song}, {Patience}, {Lafreni{\`e}re},
  {Doyon}, \& {Nielsen}}]{galicher2016}
{Galicher}, R., {Marois}, C., {Macintosh}, B., {et~al.} 2016, \aap, 594, A63

\bibitem[{{Gomez Gonzalez} {et~al.}(2018){Gomez Gonzalez}, {Absil}, \& {Van
  Droogenbroeck}}]{gomez-gonzalez2018}
{Gomez Gonzalez}, C.~A., {Absil}, O., \& {Van Droogenbroeck}, M. 2018, \aap,
  613, A71

\bibitem[{{Gomez Gonzalez} {et~al.}(2017){Gomez Gonzalez}, {Wertz}, {Absil},
  {Christiaens}, {Defr{\`e}re}, {Mawet}, {Milli}, {Absil}, {Van Droogenbroeck},
  {Cantalloube}, {Hinz}, {Skemer}, {Karlsson}, \&
  {Surdej}}]{gomez-gonzalez2017}
{Gomez Gonzalez}, C.~A., {Wertz}, O., {Absil}, O., {et~al.} 2017, \aj, 154, 7

\bibitem[{{Goodman} \& {Weare}(2010)}]{goodman2010}
{Goodman}, J. \& {Weare}, J. 2010, Communications in Applied Mathematics and
  Computational Science, Vol.~5, No.~1, p.~65-80, 2010, 5, 65

\bibitem[{{Hagelberg} {et~al.}(2016){Hagelberg}, {S{\'e}gransan}, {Udry}, \&
  {Wildi}}]{hagelberg2016}
{Hagelberg}, J., {S{\'e}gransan}, D., {Udry}, S., \& {Wildi}, F. 2016, \mnras,
  455, 2178

\bibitem[{{Hinkley} {et~al.}(2007){Hinkley}, {Oppenheimer}, {Soummer},
  {Sivaramakrishnan}, {Roberts}, {Kuhn}, {Makidon}, {Perrin}, {Lloyd},
  {Kratter}, \& {Brenner}}]{hinkley2007}
{Hinkley}, S., {Oppenheimer}, B.~R., {Soummer}, R., {et~al.} 2007, \apj, 654,
  633

\bibitem[{{Hunziker} {et~al.}(2018){Hunziker}, {Quanz}, {Amara}, \&
  {Meyer}}]{hunziker2018}
{Hunziker}, S., {Quanz}, S.~P., {Amara}, A., \& {Meyer}, M.~R. 2018, \aap, 611,
  A23

\bibitem[{{Jensen-Clem} {et~al.}(2018){Jensen-Clem}, {Mawet}, {Gomez Gonzalez},
  {Absil}, {Belikov}, {Currie}, {Kenworthy}, {Marois}, {Mazoyer}, {Ruane},
  {Tanner}, \& {Cantalloube}}]{jensen-clem2018}
{Jensen-Clem}, R., {Mawet}, D., {Gomez Gonzalez}, C.~A., {et~al.} 2018, \aj,
  155, 19

\bibitem[{{Lafreni{\`e}re} {et~al.}(2007){Lafreni{\`e}re}, {Marois}, {Doyon},
  {Nadeau}, \& {Artigau}}]{lafreniere2007}
{Lafreni{\`e}re}, D., {Marois}, C., {Doyon}, R., {Nadeau}, D., \& {Artigau},
  {\'E}. 2007, \apj, 660, 770

\bibitem[{{Lagrange} {et~al.}(2010){Lagrange}, {Bonnefoy}, {Chauvin}, {Apai},
  {Ehrenreich}, {Boccaletti}, {Gratadour}, {Rouan}, {Mouillet}, {Lacour}, \&
  {Kasper}}]{lagrange2010}
{Lagrange}, A.-M., {Bonnefoy}, M., {Chauvin}, G., {et~al.} 2010, Science, 329,
  57

\bibitem[{{Lagrange} {et~al.}(2009){Lagrange}, {Gratadour}, {Chauvin}, {Fusco},
  {Ehrenreich}, {Mouillet}, {Rousset}, {Rouan}, {Allard}, {Gendron}, {Charton},
  {Mugnier}, {Rabou}, {Montri}, \& {Lacombe}}]{lagrange2009}
{Lagrange}, A.-M., {Gratadour}, D., {Chauvin}, G., {et~al.} 2009, \aap, 493,
  L21

\bibitem[{Levenberg(1944)}]{levenberg1944}
Levenberg, K. 1944, Quarterly of Applied Mathematics, 2, 164

\bibitem[{{Macintosh} {et~al.}(2015){Macintosh}, {Graham}, {Barman}, {De Rosa},
  {Konopacky}, {Marley}, {Marois}, {Nielsen}, {Pueyo}, {Rajan}, {Rameau},
  {Saumon}, {Wang}, {Patience}, {Ammons}, {Arriaga}, {Artigau}, {Beckwith},
  {Brewster}, {Bruzzone}, {Bulger}, {Burningham}, {Burrows}, {Chen}, {Chiang},
  {Chilcote}, {Dawson}, {Dong}, {Doyon}, {Draper}, {Duch{\^e}ne}, {Esposito},
  {Fabrycky}, {Fitzgerald}, {Follette}, {Fortney}, {Gerard}, {Goodsell},
  {Greenbaum}, {Hibon}, {Hinkley}, {Cotten}, {Hung}, {Ingraham},
  {Johnson-Groh}, {Kalas}, {Lafreniere}, {Larkin}, {Lee}, {Line}, {Long},
  {Maire}, {Marchis}, {Matthews}, {Max}, {Metchev}, {Millar-Blanchaer},
  {Mittal}, {Morley}, {Morzinski}, {Murray-Clay}, {Oppenheimer}, {Palmer},
  {Patel}, {Perrin}, {Poyneer}, {Rafikov}, {Rantakyr{\"o}}, {Rice}, {Rojo},
  {Rudy}, {Ruffio}, {Ruiz}, {Sadakuni}, {Saddlemyer}, {Salama}, {Savransky},
  {Schneider}, {Sivaramakrishnan}, {Song}, {Soummer}, {Thomas}, {Vasisht},
  {Wallace}, {Ward-Duong}, {Wiktorowicz}, {Wolff}, \&
  {Zuckerman}}]{macintosh2015}
{Macintosh}, B., {Graham}, J.~R., {Barman}, T., {et~al.} 2015, Science, 350, 64

\bibitem[{{Macintosh} {et~al.}(2005){Macintosh}, {Poyneer}, {Sivaramakrishnan},
  \& {Marois}}]{macintosh2005}
{Macintosh}, B., {Poyneer}, L., {Sivaramakrishnan}, A., \& {Marois}, C. 2005,
  in \procspie, Vol. 5903, Astronomical Adaptive Optics Systems and
  Applications II, ed. R.~K. {Tyson} \& M.~{Lloyd-Hart}, 170--177

\bibitem[{{Macintosh} {et~al.}(2008){Macintosh}, {Graham}, {Palmer}, {Doyon},
  {Dunn}, {Gavel}, {Larkin}, {Oppenheimer}, {Saddlemyer}, {Sivaramakrishnan},
  {Wallace}, {Bauman}, {Erickson}, {Marois}, {Poyneer}, \&
  {Soummer}}]{macintosh2008}
{Macintosh}, B.~A., {Graham}, J.~R., {Palmer}, D.~W., {et~al.} 2008, in Society
  of Photo-Optical Instrumentation Engineers (SPIE) Conference Series, Vol.
  7015, Society of Photo-Optical Instrumentation Engineers (SPIE) Conference
  Series, 18

\bibitem[{{Marley} {et~al.}(2007){Marley}, {Fortney}, {Hubickyj},
  {Bodenheimer}, \& {Lissauer}}]{marley2007}
{Marley}, M.~S., {Fortney}, J.~J., {Hubickyj}, O., {Bodenheimer}, P., \&
  {Lissauer}, J.~J. 2007, \apj, 655, 541

\bibitem[{{Marois} {et~al.}(2014){Marois}, {Correia}, {Galicher}, {Ingraham},
  {Macintosh}, {Currie}, \& {De Rosa}}]{marois2014}
{Marois}, C., {Correia}, C., {Galicher}, R., {et~al.} 2014, in \procspie, Vol.
  9148, Adaptive Optics Systems IV, 91480U

\bibitem[{{Marois} {et~al.}(2006){Marois}, {Lafreni{\`e}re}, {Doyon},
  {Macintosh}, \& {Nadeau}}]{marois2006}
{Marois}, C., {Lafreni{\`e}re}, D., {Doyon}, R., {Macintosh}, B., \& {Nadeau},
  D. 2006, \apj, 641, 556

\bibitem[{{Marois} {et~al.}(2008){Marois}, {Macintosh}, {Barman}, {Zuckerman},
  {Song}, {Patience}, {Lafreni{\`e}re}, \& {Doyon}}]{marois2008}
{Marois}, C., {Macintosh}, B., {Barman}, T., {et~al.} 2008, Science, 322, 1348

\bibitem[{{Marois} {et~al.}(2010{\natexlab{a}}){Marois}, {Macintosh}, \&
  {V{\'e}ran}}]{marois2010a}
{Marois}, C., {Macintosh}, B., \& {V{\'e}ran}, J.-P. 2010{\natexlab{a}}, in
  \procspie, Vol. 7736, Adaptive Optics Systems II, 77361J

\bibitem[{{Marois} {et~al.}(2010{\natexlab{b}}){Marois}, {Zuckerman},
  {Konopacky}, {Macintosh}, \& {Barman}}]{marois2010b}
{Marois}, C., {Zuckerman}, B., {Konopacky}, Q.~M., {Macintosh}, B., \&
  {Barman}, T. 2010{\natexlab{b}}, \nat, 468, 1080

\bibitem[{Marquardt(1963)}]{marquardt1963}
Marquardt, D. 1963, Journal of the Society for Industrial and Applied
  Mathematics, 11, 431

\bibitem[{Mattson {et~al.}(2004)Mattson, Sanders, \& Massingill}]{mattson2004}
Mattson, T., Sanders, B., \& Massingill, B. 2004, Patterns for Parallel
  Programming, 1st edn. (Addison-Wesley Professional)

\bibitem[{{Mawet} {et~al.}(2013){Mawet}, {Absil}, {Delacroix}, {Girard},
  {Milli}, {O'Neal}, {Baudoz}, {Boccaletti}, {Bourget}, {Christiaens},
  {Forsberg}, {Gonte}, {Habraken}, {Hanot}, {Karlsson}, {Kasper}, {Lizon},
  {Muzic}, {Olivier}, {Pe{\~n}a}, {Slusarenko}, {Tacconi-Garman}, \&
  {Surdej}}]{mawet2013}
{Mawet}, D., {Absil}, O., {Delacroix}, C., {et~al.} 2013, \aap, 552, L13

\bibitem[{{Mawet} {et~al.}(2014){Mawet}, {Milli}, {Wahhaj}, {Pelat}, {Absil},
  {Delacroix}, {Boccaletti}, {Kasper}, {Kenworthy}, {Marois}, {Mennesson}, \&
  {Pueyo}}]{mawet2014}
{Mawet}, D., {Milli}, J., {Wahhaj}, Z., {et~al.} 2014, \apj, 792, 97

\bibitem[{{Mawet} {et~al.}(2005){Mawet}, {Riaud}, {Absil}, \&
  {Surdej}}]{mawet2005}
{Mawet}, D., {Riaud}, P., {Absil}, O., \& {Surdej}, J. 2005, \apj, 633, 1191

\bibitem[{{Morzinski} {et~al.}(2015){Morzinski}, {Males}, {Skemer}, {Close},
  {Hinz}, {Rodigas}, {Puglisi}, {Esposito}, {Riccardi}, {Pinna}, {Xompero},
  {Briguglio}, {Bailey}, {Follette}, {Kopon}, {Weinberger}, \&
  {Wu}}]{morzinski2015}
{Morzinski}, K.~M., {Males}, J.~R., {Skemer}, A.~J., {et~al.} 2015, \apj, 815,
  108

\bibitem[{Nelder \& Mead(1965)}]{nelder1965}
Nelder, J.~A. \& Mead, R. 1965, The Computer Journal, 7, 308

\bibitem[{{Oppenheimer} \& {Hinkley}(2009)}]{oppenheimer2009}
{Oppenheimer}, B.~R. \& {Hinkley}, S. 2009, \araa, 47, 253

\bibitem[{{Pueyo} {et~al.}(2012){Pueyo}, {Crepp}, {Vasisht}, {Brenner},
  {Oppenheimer}, {Zimmerman}, {Hinkley}, {Parry}, {Beichman}, {Hillenbrand},
  {Roberts}, {Dekany}, {Shao}, {Burruss}, {Bouchez}, {Roberts}, \&
  {Soummer}}]{pueyo2012}
{Pueyo}, L., {Crepp}, J.~R., {Vasisht}, G., {et~al.} 2012, \apjs, 199, 6

\bibitem[{{Rajan} {et~al.}(2017){Rajan}, {Rameau}, {De Rosa}, {Marley},
  {Graham}, {Macintosh}, {Marois}, {Morley}, {Patience}, {Pueyo}, {Saumon},
  {Ward-Duong}, {Ammons}, {Arriaga}, {Bailey}, {Barman}, {Bulger}, {Burrows},
  {Chilcote}, {Cotten}, {Czekala}, {Doyon}, {Duch{\^e}ne}, {Esposito},
  {Fitzgerald}, {Follette}, {Fortney}, {Goodsell}, {Greenbaum}, {Hibon},
  {Hung}, {Ingraham}, {Johnson-Groh}, {Kalas}, {Konopacky}, {Lafreni{\`e}re},
  {Larkin}, {Maire}, {Marchis}, {Metchev}, {Millar-Blanchaer}, {Morzinski},
  {Nielsen}, {Oppenheimer}, {Palmer}, {Patel}, {Perrin}, {Poyneer},
  {Rantakyr{\"o}}, {Ruffio}, {Savransky}, {Schneider}, {Sivaramakrishnan},
  {Song}, {Soummer}, {Thomas}, {Vasisht}, {Wallace}, {Wang}, {Wiktorowicz}, \&
  {Wolff}}]{rajan2017}
{Rajan}, A., {Rameau}, J., {De Rosa}, R.~J., {et~al.} 2017, \aj, 154, 10

\bibitem[{{Rameau} {et~al.}(2013{\natexlab{a}}){Rameau}, {Chauvin}, {Lagrange},
  {Boccaletti}, {Quanz}, {Bonnefoy}, {Girard}, {Delorme}, {Desidera}, {Klahr},
  {Mordasini}, {Dumas}, \& {Bonavita}}]{rameau2013a}
{Rameau}, J., {Chauvin}, G., {Lagrange}, A.-M., {et~al.} 2013{\natexlab{a}},
  \apjl, 772, L15

\bibitem[{{Rameau} {et~al.}(2013{\natexlab{b}}){Rameau}, {Chauvin}, {Lagrange},
  {Klahr}, {Bonnefoy}, {Mordasini}, {Bonavita}, {Desidera}, {Dumas}, \&
  {Girard}}]{rameau2013}
{Rameau}, J., {Chauvin}, G., {Lagrange}, A.-M., {et~al.} 2013{\natexlab{b}},
  \aap, 553, A60

\bibitem[{{Rameau} {et~al.}(2013{\natexlab{c}}){Rameau}, {Chauvin}, {Lagrange},
  {Meshkat}, {Boccaletti}, {Quanz}, {Currie}, {Mawet}, {Girard}, {Bonnefoy}, \&
  {Kenworthy}}]{rameau2013b}
{Rameau}, J., {Chauvin}, G., {Lagrange}, A.-M., {et~al.} 2013{\natexlab{c}},
  \apjl, 779, L26

\bibitem[{{Ruffio} {et~al.}(2018){Ruffio}, {Mawet}, {Czekala}, {Macintosh}, {De
  Rosa}, {Ruane}, {Bottom}, {Pueyo}, {Wang}, {Hirsch}, {Zhu}, \&
  {Nielsen}}]{ruffio2018}
{Ruffio}, J.-B., {Mawet}, D., {Czekala}, I., {et~al.} 2018, \aj, 156, 196

\bibitem[{{Soummer} {et~al.}(2012){Soummer}, {Pueyo}, \&
  {Larkin}}]{soummer2012}
{Soummer}, R., {Pueyo}, L., \& {Larkin}, J. 2012, \apjl, 755, L28

\bibitem[{{Wahhaj} {et~al.}(2015){Wahhaj}, {Cieza}, {Mawet}, {Yang}, {Canovas},
  {de Boer}, {Casassus}, {M{\'e}nard}, {Schreiber}, {Liu}, {Biller}, {Nielsen},
  \& {Hayward}}]{wahhaj2015}
{Wahhaj}, Z., {Cieza}, L.~A., {Mawet}, D., {et~al.} 2015, \aap, 581, A24

\bibitem[{{Wang} {et~al.}(2018){Wang}, {Perrin}, {Savransky}, {Arriaga},
  {Chilcote}, {De Rosa}, {Millar-Blanchaer}, {Marois}, {Rameau}, {Wolff},
  {Shapiro}, {Ruffio}, {Maire}, {Marchis}, {Graham}, {Macintosh}, {Ammons},
  {Bailey}, {Barman}, {Bruzzone}, {Bulger}, {Cotten}, {Doyon}, {Duch{\^e}ne},
  {Fitzgerald}, {Follette}, {Goodsell}, {Greenbaum}, {Hibon}, {Hung},
  {Ingraham}, {Kalas}, {Konopacky}, {Larkin}, {Marley}, {Metchev}, {Nielsen},
  {Oppenheimer}, {Palmer}, {Patience}, {Poyneer}, {Pueyo}, {Rajan},
  {Rantakyr{\"o}}, {Schneider}, {Sivaramakrishnan}, {Song}, {Soummer},
  {Thomas}, {Wallace}, {Ward-Duong}, \& {Wiktorowicz}}]{wang2018}
{Wang}, J.~J., {Perrin}, M.~D., {Savransky}, D., {et~al.} 2018, Journal of
  Astronomical Telescopes, Instruments, and Systems, 4, 018002

\bibitem[{{Wertz} {et~al.}(2017){Wertz}, {Absil}, {G{\'o}mez Gonz{\'a}lez},
  {Milli}, {Girard}, {Mawet}, \& {Pueyo}}]{wertz2017}
{Wertz}, O., {Absil}, O., {G{\'o}mez Gonz{\'a}lez}, C.~A., {et~al.} 2017, \aap,
  598, A83

\bibitem[{{Zimmerman} {et~al.}(2011){Zimmerman}, {Brenner}, {Oppenheimer},
  {Parry}, {Hinkley}, {Hunt}, \& {Roberts}}]{zimmerman2011}
{Zimmerman}, N., {Brenner}, D., {Oppenheimer}, B.~R., {et~al.} 2011, \pasp,
  123, 746

\end{thebibliography}

\end{document}